%% file: AUMic_Sezestre.tex
\documentclass[bibyear]{aa}

\usepackage{graphicx}
\usepackage{txfonts}
\usepackage{natbib}
\bibpunct{(}{)}{;}{a}{}{,}
\usepackage{scrextend}			
\usepackage{tikz}
\usepackage{caption,subcaption}
\usepackage{multirow}
\usepackage{hyperref}
\usepackage{gensymb}
\definecolor{Acol}{RGB}{7,32,244}
\definecolor{Bcol}{RGB}{255,0,23}
\definecolor{Ccol}{RGB}{0,127,36}
\definecolor{Dcol}{RGB}{0,192,191}
\definecolor{Ecol}{RGB}{196,14,184}

\begin{document} 

\newcommand{\be}{$\beta$}
\newcommand{\R}{$R_0$}
\newcommand{\XX}{$\chi^2_r$}

\newcommand{\ma}[1]{\mathrm{#1}}
\newcommand{\uma}[1]{^{\mathrm{#1}}}
\newcommand{\dma}[1]{_{\mathrm{#1}}}

\title{Expelled grains from an unseen parent body around AU Mic}
\titlerunning{Expelled grains from an unseen parent body around AU Mic}

   \author{\'E. Sezestre\inst{1}
          \and
          J.-C. Augereau\inst{1}
          \and
          A. Boccaletti\inst{2}
          \and
          P. Th\'ebault\inst{2}
          }

   \institute{Univ. Grenoble Alpes, CNRS, IPAG, F-38000 Grenoble, France 
     \and
     LESIA, Observatoire de Paris, CNRS, Universit\'e Paris Diderot,
     Universit\'e Pierre et Marie Curie, 5 place Jules Janssen, 92190
     Meudon, France
\\ correspondence: \texttt{elie.sezestre@univ-grenoble-alpes.fr}
   }

   \date{Received 28/04/2017; accepted 11/09/2017}

 
  \abstract
   {
   Recent observations of the edge-on debris disk of \object{AU Mic} have
     revealed asymmetric, fast outward-moving arch-like structures above the disk midplane. 
     Although asymmetries are frequent in debris disks,
     no model can readily explain the characteristics of these features.}
   {
   We present a model aiming to reproduce the dynamics
     of these structures, more specifically their high projected speeds and their apparent position. 
     We test the hypothesis of dust emitted by a point source
    and then expelled from the system by the strong stellar wind of this young, M-type star.
    In this model, we make the assumption that the dust grains follow the same dynamics as the structures,
     i.e. they are not local density enhancements. }
   {
   We perform numerical simulations of test particle trajectories to
     explore the available parameter space, in particular the radial location $R_{0}$ 
     of the dust producing parent body and the size of the dust grains 
     as parameterized by the value of $\beta$ (ratio of stellar wind and radiation pressure forces over gravitation). 
     We consider both the case of a static and an orbiting parent body.}
   {
   We find that, for all considered scenarii (static or moving parent body), 
   there is always a set of ($R_0, \beta$) parameters able to fit the observed features. 
   The common characteristics of these solutions is that they all require a high value of $\beta$, of around 6. 
   This means that the star is probably very active and the grains composing the structures are sub-micronic,
   in order for observable grains to reach such high $\beta$ values. 
   As for the location of the hypothetical parent body, we  find it to be closer-in than the planetesimal belt,  
   around $8 \pm 2$~au (orbiting case) or $28 \pm 7$~au (static case). 
   A nearly periodic process of dust emission appears, of 2 years in the orbiting scenarii, and 7 years in the static case. }
   {
   We show that the scenario of sequential dust releases by an unseen,  point-source  parent body is able to explain the radial behaviour of the observed structures. 
   We predict the evolution of the structures 
   to help future observations 
   to discriminate between the different parent body configurations that have been considered.
    In the orbiting parent body scenario,  we expect new structures to appear on the northwest side of the disk in the coming years.
}


   \maketitle


\section{Introduction}
\label{sec:intro}

AU Mic is an active M-type star, in the $\beta$ Pictoris moving group,
with an age of 23 $\pm$ 3 Myr \citep{Mamajek2014}.  Its debris disk,
seen almost edge-on, has been imaged for the first time by
\cite{Kalas2004} in the optical.  The dust seen in scattered light has
been shown to originate from collisional grinding of planetesimals
arranged in a belt at $\sim$35-40~au
\citep{Augereau2006,Strubbe2006,Schuppler2015}.  The belt was later
resolved by millimeter imaging \citep{Wilner2012,Macgregor2013}.
Because the star is very active, the dynamics of the dust grains is
believed to be strongly affected by the stellar wind.

\begin{figure*}
\begin{center}
\begin{tikzpicture}
\node at (0,0) {\includegraphics[height=0.9\textwidth, angle=90, trim=7cm 15mm 6cm 15mm]{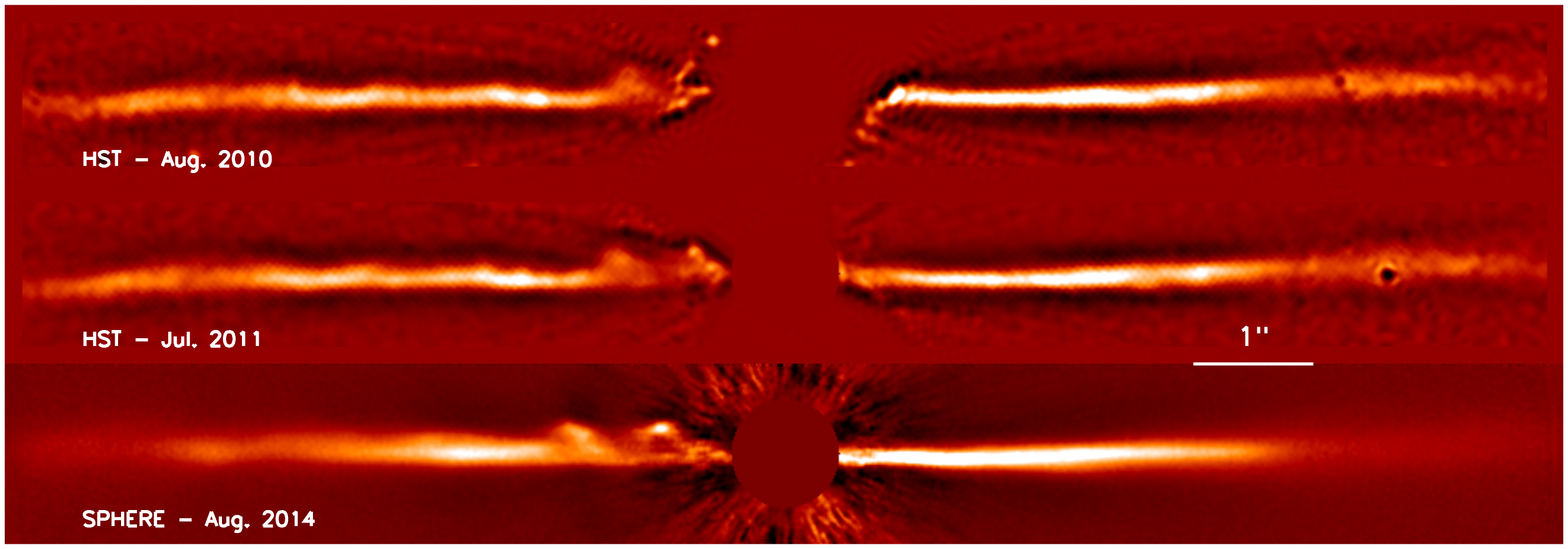} };
\node at (-1.5,-1.5) {\small \color{white}{A}} ;
\node at (-2.5,-1.5) {\small \color{white}{B}} ;
\node at (-4.0,-1.5) {\small \color{white}{C}} ;
\node at (-5.5,-1.5) {\small \color{white}{D}} ;
\node at (-7.3,-1.5) {\small \color{white}{E}} ;

\node at (-1.3,-1.7) {\small \color{white}{|}} ; 
\node at (-2.3,-1.7) {\small \color{white}{|}} ;
\node at (-3.8,-1.7) {\small \color{white}{|}} ;
\node at (-5.3,-1.7) {\small \color{white}{|}} ;
\node at (-7.1,-1.7) {\small \color{white}{|}} ;

\node at (-0.95,0.2) {\small \color{white}{|}} ; 
\node at (-1.8,0.2) {\small \color{white}{|}} ;
\node at (-3.2,0.2) {\small \color{white}{|}} ;
\node at (-4.4,0.2) {\small \color{white}{|}} ;
\node at (-6.,0.2) {\small \color{white}{|}} ;

\node at (-1.7,2.1) {\small \color{white}{|}} ;
\node at (-3.,2.1) {\small \color{white}{|}} ;
\node at (-4.2,2.1) {\small \color{white}{|}} ;
\node at (-5.8,2.1) {\small \color{white}{|}} ;
\end{tikzpicture}
\end{center}
\caption{2010 and 2011 HST/STIS and 2014 VLT/SPHERE images of the debris disk of AU Mic. 
The five structures are identified in the bottom panel.}
\label{imgBoc}
\end{figure*}

Recently, \cite{Boccaletti2015} have revealed five fast-moving,
arch-like, vertical features in this disk in scattered light imaging
with HST/STIS \citep{Schneider2014} and 
VLT/SPHERE \citep{Beuzit2008} at three different epochs. 
The five structures, named A to E, are shown in Figure~\ref{imgBoc}.  
They are identified in the 2010, 2011 and 2014 images, except for the A
structure which was too close to the star in 2010.  Curiously, these
structures are all located on one single side of the disk and they all
show an outward migration.  For the D and E structures, the velocities
are such that these features could match asymmetries identified in
earlier, multiple wavelength observations \citep{Liu2004, Krist2005,
  Metchev2005, Fitzgerald2007}. Although they move outward, the
arch-like structures seem stable in shape over a time span of a few
years.

\begin{figure}[tb]
\includegraphics[width=0.49\textwidth, height=!,trim=0 0.5cm 0 1cm]{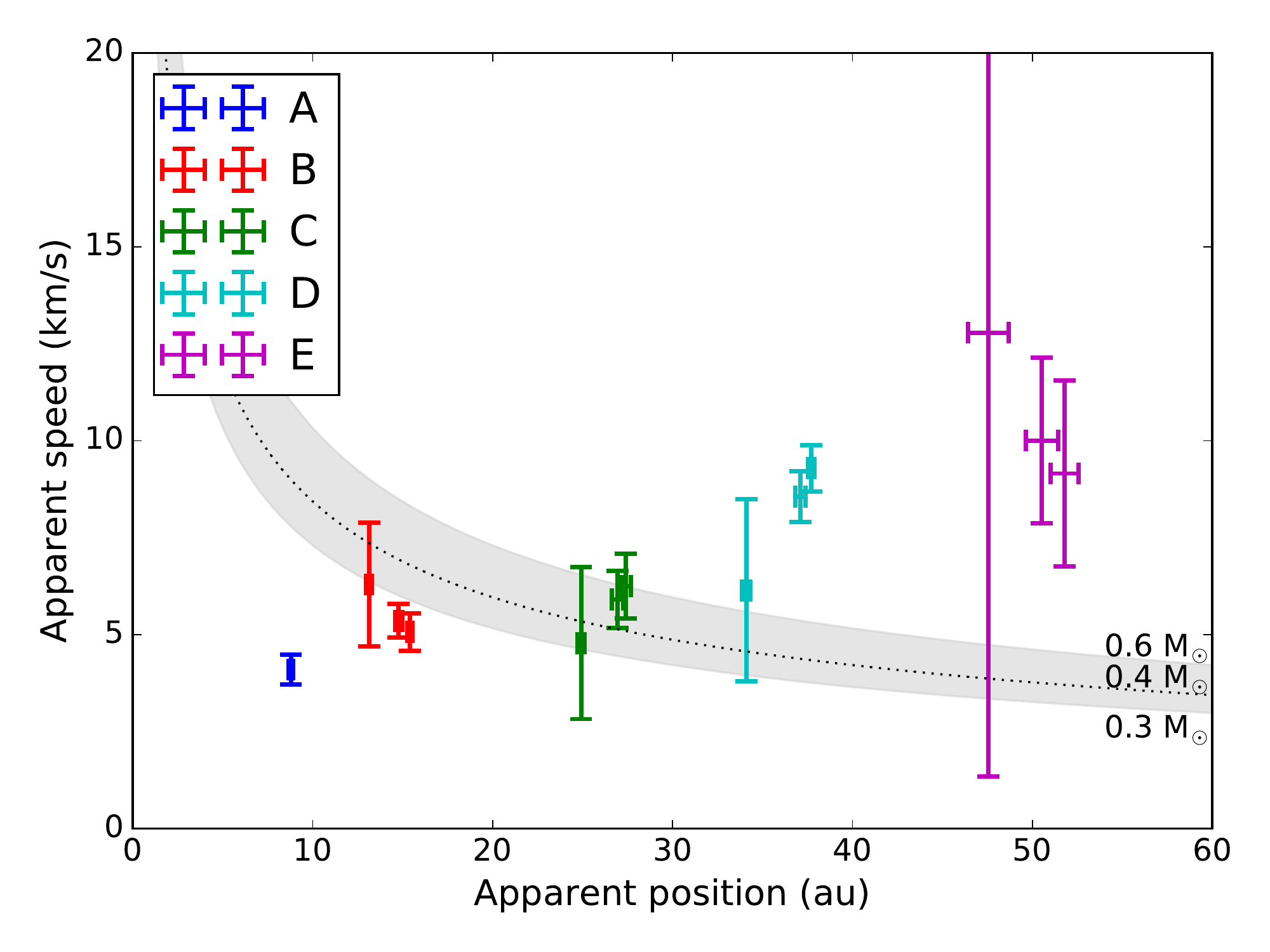}
\caption{Apparent speeds of the structures in the AU~Mic debris  disk 
  derived from the observations (Tabs.~\ref{tabPos} and
  \ref{tabValVit}). The gray region shows the escape velocities for
  stellar masses ranging from 0.3 to 0.6~M$_{\odot}$, the dotted line
  corresponding to a mass of 0.4~M$_{\odot}$.}
\label{figVitObs}
\end{figure}

The projected speeds derived from the observations are displayed in
Figure~\ref{figVitObs}. One can see that they increase with rising
distance to the star. The two outermost, at least, are exceeding the
local escape velocity. There is currently no theorical framework to
readily explain this behaviour \citep[see e.g.][ for a recent review
  on debris disks]{Matthews2014}. Any dynamical process involving
copious amount of gas, such as radiation-driven disk winds which may
allow to reach high velocities are excluded by the low amount of gas
remaining around AU Mic (\citeauthor{Roberge2005}
\citeyear{Roberge2005} fixed an upper limit of H$_2$ mass at
0.07~$M_{\oplus}$).  
Vertical resonances with a planetary companion can form arch-like structures, 
but they stay on a Keplerian orbit 
\citep[analogous to the case of Saturn's moons, see e.g.][]{Weiss2009}.
Lindblad resonances can induce spiral density arms phase-locked with a perturber, 
but if each structure corresponds to an arm, 
they would be observed on both sides of the disk.
Concentric eccentric rings resulting from massive collisions of asteroid-like
objects can produce local intensity maxima \citep[e.g.][]{Kral2015},
but this process requires a time scale of 100 years, while the
structures escape the system in tens of years.

 In this study, we aim to reproduce the observed high speeds 
and apparent positions of the structures. \rm
We leave aside in this paper the
origin of the vertical elevation of the structures. In
Section~\ref{secModel} we  describe  our model. There, we assume that the
NW/SE asymmetry can be explained by a local process of dust release.
This hypothetic emission source will be refered to  as  \textit{parent
  body} in the following, without further
specification. \citet{Boccaletti2015} , for instance,  proposed that this
could correspond to a planet, which magnetosphere or dust
circumplanetary ring would be interacting with the stellar wind.
The dust released by the parent body  is exposed to the stellar wind. 
The resulting wind pressure can put this dust on unbound trajectories,
 achieving the observed high projected speeds. 
In Section~\ref{secResults}, we explore the case of a static parent body,
that would for example mimic a source of dust due to a giant collision
\citep[e.g.][]{Jackson2014, Kral2015}  or a localised dust avalanche \citep{Chiang2017}, 
and the case of an orbiting
parent body, for example, a young planet. 
 However, we emphasize that we do not suppose any specific dust production process in our study.
Instead, we focus on the dynamical evolution of the dust right after its release, 
and any dust production mechanism will have to comply 
with the constraints we derive on the dust properties and dynamics. 
We discuss our findings, the
influence of the parameters, and the implications in
Section~\ref{secDiscuss}.


\section{Model}
\label{secModel}

We develop a model that aims to investigate the dynamics of dust
particles released by a singular parent body and affected by a strong
stellar wind pressure force. Throughout this paper, we make the
important assumption that the observed displacements correspond to the
actual proper motion of the particles, and not to a wave pattern,
which implies that the particles are supposed to have the same
projected speeds as the observed structures. We seek to discuss if
this assumption could yield to scenarii that allow to reproduce the
observed speeds in the AU~Mic debris disk, and which conditions must
be fulfilled.

\subsection{Parent body}
\label{subsecParent}

To break the symmetry of the disk, we need an asymmetric process of
dust production.  The dust arranged in the fast-moving structures is
thus assumed to be locally released by an unresolved, unknown source: the  "parent body".  
This hypothetical parent body will not be described further, but we note
that it must be massive enough to produce a significant amount of
dust, although it should remain faint enough to be undetected with
current instrumentation.  
Actually, the upper mass limit is fixed by the non detection of point source in SPHERE imaging.  
This implies a compact parent body smaller than 6~Jupiter masses beyond 10~au 
\citep[see Methods of][]{Boccaletti2015}.
The process of dust production is neither
described in this model, except that it must be sporadic otherwise we
would observe one single, continuous feature.
 We can thus exclude the flares of the star for being the trigger events responsible for the arch
formation as they are too frequent \citep[0.9 flare per hour following][]{Kunkel1973}. 

The parent body is assumed to be at a distance~\R\ from the star in
the plane of the main disk. When supposed to be revolving around the
star, its orbit is considered to be circular.
The dust particles are released with the local Keplerian speed.

\subsection{Pressure forces}
\label{subsecPressureForce}

\begin{figure}
\centering
\includegraphics[width=0.49\textwidth, height=!,trim=0 0.7cm 0 0.5cm,clip]{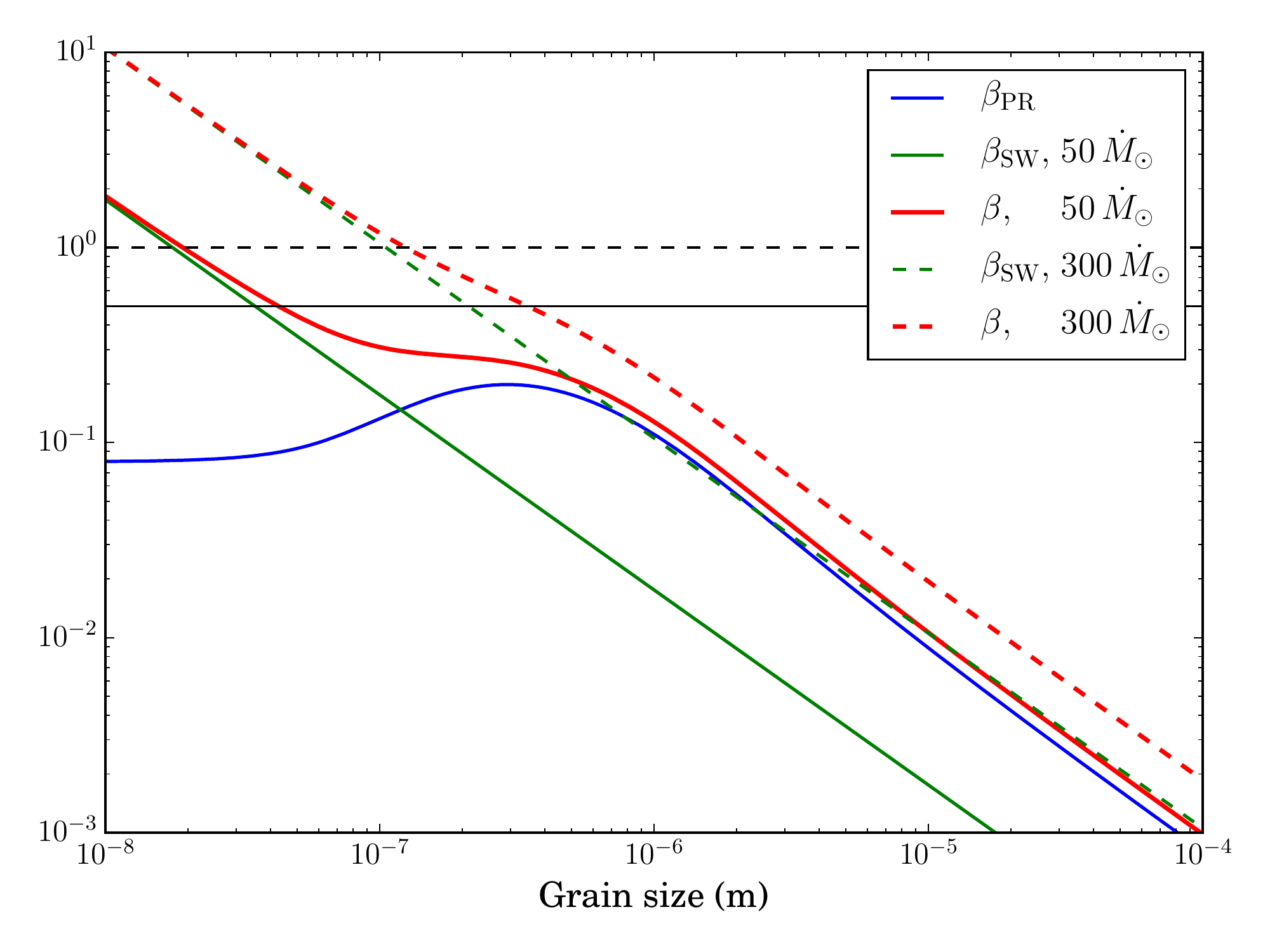}
\caption{ \be\ values as a function of grain size.
The material used for $\beta\dma{PR}$ is M1 of \cite{Schuppler2015}. 
Two hypothesis for the mass-loss rate of the star are shown (solid and dashed lines).
The horizontal solid black line is the upper limit for bound trajectories 
assuming zero eccentricity for the parent body, 
and the dashed line is the limit between "normal" and "abnormal" parabolic trajectories 
(see Sec.~\ref{subsecBehaviour}). }
\label{figBetaSize}
\end{figure}

Supposing that the observed velocities correspond to the effective
speeds of the particles, that means they accelerate outwards once
released, to finally exceed the escape velocity.  
As the particle size decreases, 
they are more affected by the pressure coming from the stellar wind.  
The radiation pressure is also more efficient 
although it remains low because AU Mic is an M-type star.  
These processes can accelerate outward the particles, 
provided the total pressure force exceeds the gravitational force.

The two pressure forces will be described by a single parameter noted
$\beta$ in the following.  It is the ratio of the wind plus radiation
forces to the gravitational force
\begin{equation}
\beta = \frac{|| \mathbf{F}_{\rm SW} || + || \mathbf{F}_{\rm PR}  ||}{ || \mathbf{F}_{\rm grav} || } = \beta_{\rm SW} + \beta_{\rm PR} ,
\end{equation}
under the assumption that the grain velocity~$v$ is such that $v\ll c$
and $v \ll V_{\rm sw}$ where $c$ is the light speed in vacuum and $V\dma{sw}$
is the wind speed.  $\mathbf{F}_{\rm SW}$ is the force exerted by the
stellar wind on the particle.  $\mathbf{F}_{\rm PR}$ is the radiation
force, taking into account the radiation pressure and the
Poynting-Robertson drag.  $\mathbf{F}_{\rm grav}$ is the gravitational
force of the star.  For typical silicate submicrometer-sized grains,
$\beta$ ranges from $\sim 10^{-1}$ to a few tens 
(see Fig.~1 of \citeauthor{Schuppler2015}~\citeyear{Schuppler2015} and Fig.~11 of \citeauthor{Augereau2006}~\citeyear{Augereau2006}).

The two contributions to $\beta$ can be estimated with, for example,
Eq.~28 of \cite{Augereau2006} and Eq.~6 of \cite{Strubbe2006},
respectively:
\begin{eqnarray}
\beta_{\rm SW} & = & \frac{3}{32 \pi} \frac{\dot{M}_{\star} V\dma{sw}
  C\dma{D}}{G M_{\star} \rho s } , \label{eqBsw} \\
\beta_{\rm PR} & = & \frac{3}{16 \pi} \frac{L_{\star} \langle Q\dma{PR}
  \rangle}{c \, G M_{\star} \rho s } , \mathrm{\,\,\, with \,\,\,}
\langle Q\dma{PR} \rangle = \dfrac{ \int_{\lambda} F_{\lambda}
  Q\dma{PR} \ma{d}\lambda }{ \int_{\lambda} F_{\lambda} \ma{d}\lambda
} ,
\end{eqnarray}
where $\dot{M}_{\star}$ is the stellar mass loss rate, $C\dma{D}$ the
dimensionless free molecular drag coefficient which has a value close
to 2, $G$ the gravitational constant, $M_{\star}$ the mass of the
star, $\rho$ the grain volumetric mass density, $s$ the grain radius,
$L_{\star}$  the stellar luminosity, \rm $Q\dma{PR}$ the dimensionless
radiation pressure efficiency (that depends on the grain size,
composition and wavelength), and $F_{\lambda}$ the stellar flux at
wavelength $\lambda$.

$\beta_{\rm PR}$ is independent on the grain's distance to the
star~($r$), but $\beta_{\rm SW}$ can slightly depend on~$r$
\citep[e.g. Fig. 11 in][]{Augereau2006}.  In this study, we will
neglect this effect.  $\beta$ is highly size-dependent.  For
sufficiently large grains ($s\gtrsim 1 \mu$m in the case of AU Mic),
$\beta$ varies as $s^{-1}$ \citep{Schuppler2015}.  
For smaller grain sizes, the relationship
between $\beta$ and $s$ is more complex \citep[e.g. Fig.~1
in][]{Schuppler2015} and depends both on the grain composition, 
the stellar mass loss rate $\dot{M}_{\star}$ and 
the stellar wind speed $V\dma{sw}$.
 With $\dot{M}_{\star} = 50 \, \dot{M}_{\odot}$ and $V\dma{sw} = V\dma{sw,\odot}$, 
the blowout size (grains with $\beta = 0.5$, assuming zero eccentricity for the parent body) 
is 0.04~$\mu$m (Fig.~\ref{figBetaSize}). 
This size jumps to 0.35~$\mu$m if the stellar mass-loss rate is increased to $300 \, \dot{M}_{\odot}$. 
These values are consistent with those reported in Tab.~2 of \cite{Schuppler2015} 
although they slightly differ because of minor differences in the assumed stellar properties.

\subsection{Particles behaviour}
\label{subsecBehaviour}

\begin{figure}[tp!]
\includegraphics[width=0.49\textwidth, height=!,trim=0 0 0 0]{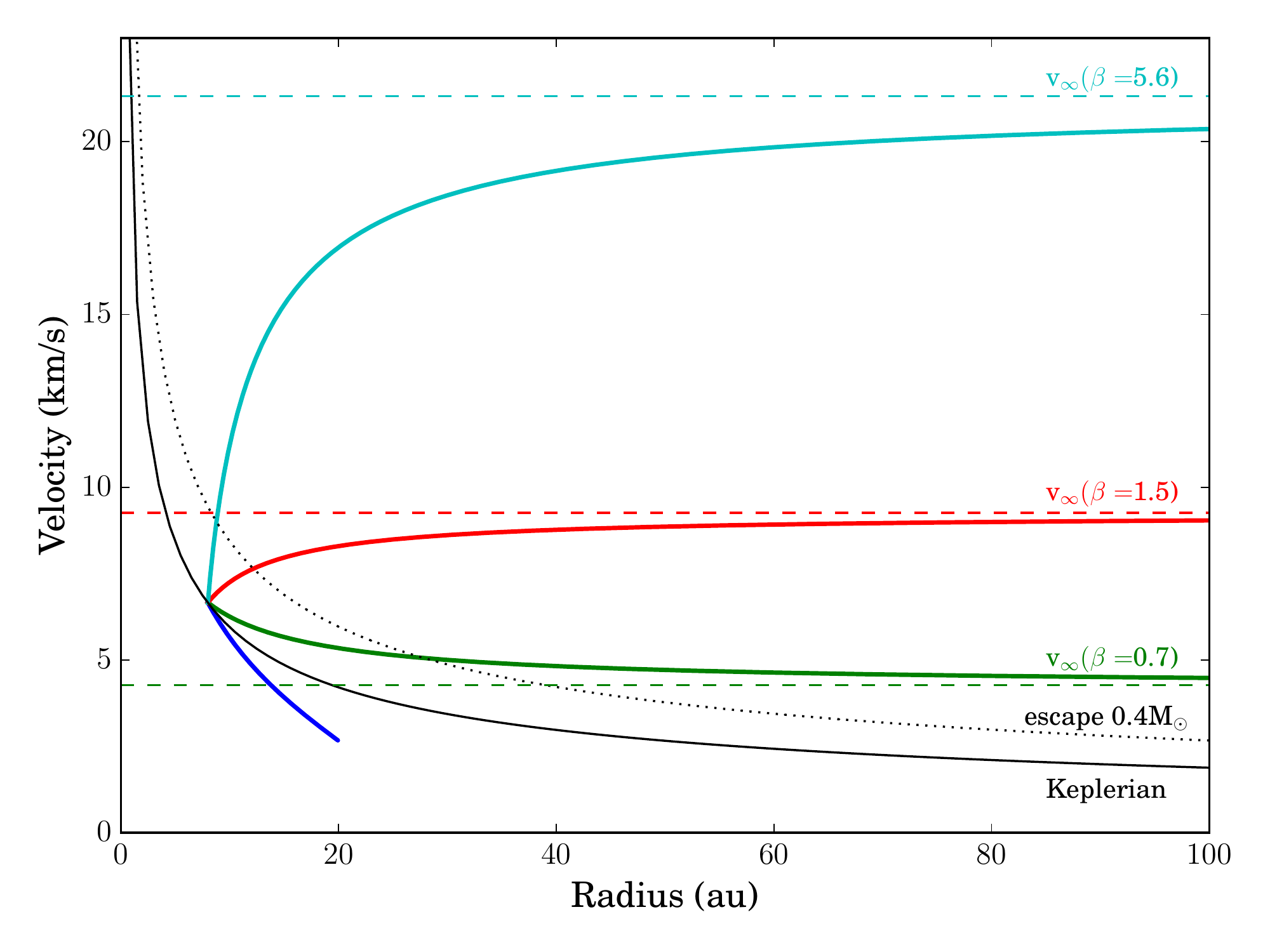}
\caption{Module of the velocity as a function of the distance from the
  star for a particle released with the Keplerian speed at  8~au . The
  solid curves correspond to $\beta$=0.3 (blue), 0.7 (green), 1.5
  (red)  and 5.6 (cyan).  The horizontal dashed lines are the asymptotic values of the
  velocity derived from Eq.~\ref{eqnVitLim}.}
\label{figVitMod}
\end{figure}

The trajectory of a grain released from a parent body strongly depends
on the $\beta$ value.  For a parent body on a circular orbit, the
released $0<\beta<0.5$ dust particles remain on bound orbits, with
eccentricities increasing with $\beta$, while the $0.5<\beta<1$
particles are placed on parabolic orbits.  Dust particles with
$\beta>1$ will, on the other hand, follow unbound, "abnormal"
parabolic trajectories, as described for example in \cite{Krivov2006}.
These $\beta>1$ grains are of particular interest in the context of
the AU Mic debris disk because their velocity continuously increases
while moving outwards, until it reaches an asymptotic value that can
be evaluated by considering the total energy per unit of mass of the
particle at a distance $r$ from the star:
\begin{equation}
e_m  = \frac{1}{2} v^2 - \frac{G M_{\star}}{r} (1-\beta) , \label{eqnEner}
\end{equation}
where~$M_{\star} (1-\beta)$ is the apparent mass of the star. The
particle is supposed to be released with the Keplerian velocity $v_0 =
\sqrt{G M_{\star} / R_0}$ at radius \R . Evaluating Eq.~\ref{eqnEner}
in \R\ thus yields:
\begin{equation}
e_m  = \frac{G M_{\star}}{2 R_0} (2\beta - 1) \, .
\end{equation}
Therefore, the asymptotic speed reached by the dust particle far away
from the star (valid for $\beta>0.5$) is given by:
\begin{equation}
v_{\infty} = v(r \rightarrow \infty) = \sqrt{(2\beta - 1)
  \frac{GM_{\star}}{R_0}} = v_0 \sqrt{2 \beta -1} . \label{eqnVitLim}
\end{equation}
For the $0.5<\beta<1$ grains, $v_{\infty}$ is smaller than $v_0$. In
this case, the asymptotic value of the velocity is reached by upper
values and the speed decreases with the distance from the star.  The
$\beta~>~1$ grains, on the other hand, reach the asymptotic value of
the velocity by lower values, and $v$ increases with $r$. This
behaviour is illustrated in Fig.~\ref{figVitMod}. As shown in
Fig.~\ref{figVitObs}, the observed apparent speeds are not compatible
with bound orbits at least for the structures D and E. An unbound
orbit is equivalent to $\beta>0.5$ in the model. Furthermore, the
global trend of increasing velocity with the distance to the star is
only reproduced by trajectories with $\beta>1$ (see
Fig.~\ref{figVitMod}).

\begin{table}[tp!]
\caption{Documented stellar parameters of \object{AU Microscopii}
  (\object{GJ803}, \object{HD~197481})}
\label{tabCaracAUMic}
\begin{tabular}{lll}
\hline
Parameter		& Value			& Reference	 \\ 
\hline
Spectral type					& M1Ve	& 	\citet{Torres2006}	\\
Age							& $23 \pm 3$ Myr 	& 	\citet{Mamajek2014}	\\ 
Distance					& 9.94 $\pm$ 0.13 pc & \citet{Perryman1997a} 		\\ 
Mass ($M_{\star}$) 				&0.3-0.6 M$_{\odot}$	& \citet{Schuppler2015}		\\
Wind speed ($V\dma{sw}$) & $4.5 \times 10^5$ m/s & \citet{Strubbe2006} \\ 
\hline
\end{tabular}
\end{table}

The strength of the pressure forces on the grains, characterized by
$\beta$, and the released position of the grains, \R, are two key
parameters in this model, and some constraints on their values and
relationship can be anticipated. For instance, should the grains be on
bound orbits ($\beta < 0.5$), their apoastron $r_a = (1-2 \beta)^{-1}
R_0$ should be sufficiently large for the particles to reach the
position of the furthest structures, around 50~au in projection
(structure E). Noting $r_E$ the apparent position of the E
structure, the condition $r_a \gtrsim r_E$ yields a strict lower
limit on $\beta$:
\begin{equation}
\beta > \frac{r_E - R_0}{2 r_E} .
\label{eqnNoSol}
\end{equation}
Nevertheless, we anticipate unbound orbits with high $\beta$ values to
best fit the observed speeds, and a power law linking $\beta$ and
\R\ can be approximated analytically. The trajectories of grains with
$\beta$ values much larger than 1 are almost radial and the limit
speed reached by the particle is given by Eq.~\ref{eqnVitLim}. The
data points to reproduce are apparent speeds at projected
distances. Let us take the  pair  ($r_D, v_D$) for the D structure as
an example, and note $\alpha$ the angle between the observer and the direction of
propagation of the particle. 
For a given projected
distance $r_D$, the greater the released distance to the star \R, the
smaller the $\alpha$ angle. In a simple approximation, we can write
that $\sin \alpha = r_D/(x R_0)$ by considering the right triangle
where $r_D$ is the side opposed to the angle $\alpha$ and assuming the
hypothenuse is $x$ times \R. Using Eq.~\ref{eqnVitLim} and the above
approximation, the apparent speed writes:
\begin{equation}
v_D \simeq v_{\infty} \, \sin \alpha \propto v_0 \sqrt{2 \beta -1} \cdot \frac{r_D}{R_0} \propto R_0^{-3/2} \sqrt{2 \beta - 1} .
\end{equation}
Therefore, we expect  that, for a given observed velocity, 
the best fits solutions  will  obey the following
relationship between $\beta$ and \R :
\begin{equation}
(2 \beta - 1) \propto R_0^3 , \label{eqnAnaBR0}
\end{equation}
 as displayed as black lines in Figs.~\ref{figSpeAdjABCDE}a, \ref{figMapEcc} and \ref{figPosMoy}a.

\begin{figure*}[tp!]
\centering
\hbox to \textwidth
{
\parbox{0.48\textwidth}{
\includegraphics[width=0.4\textwidth, height=!]{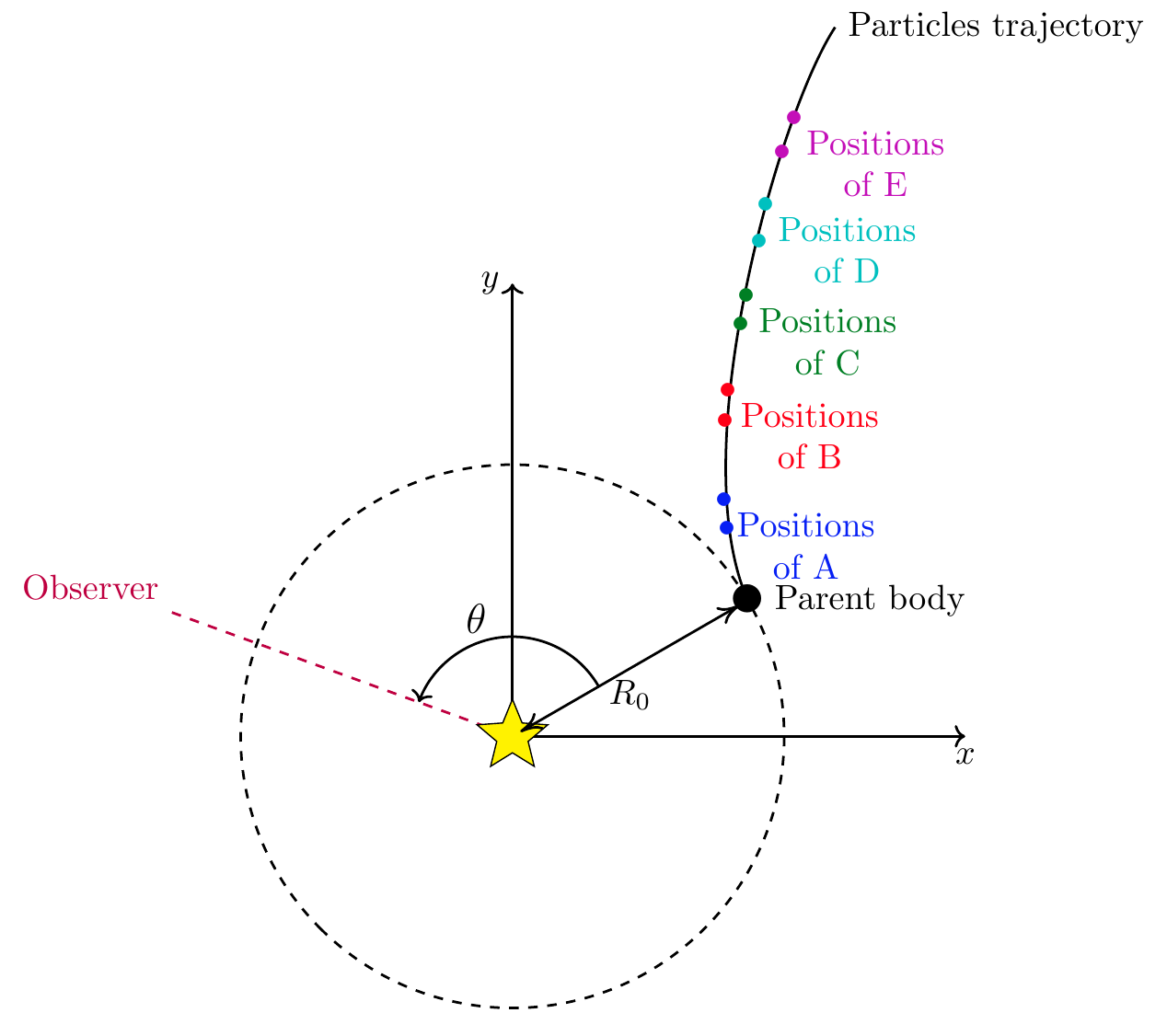}
}
\parbox{0.52\textwidth}{
\includegraphics[width=0.52\textwidth, height=!]{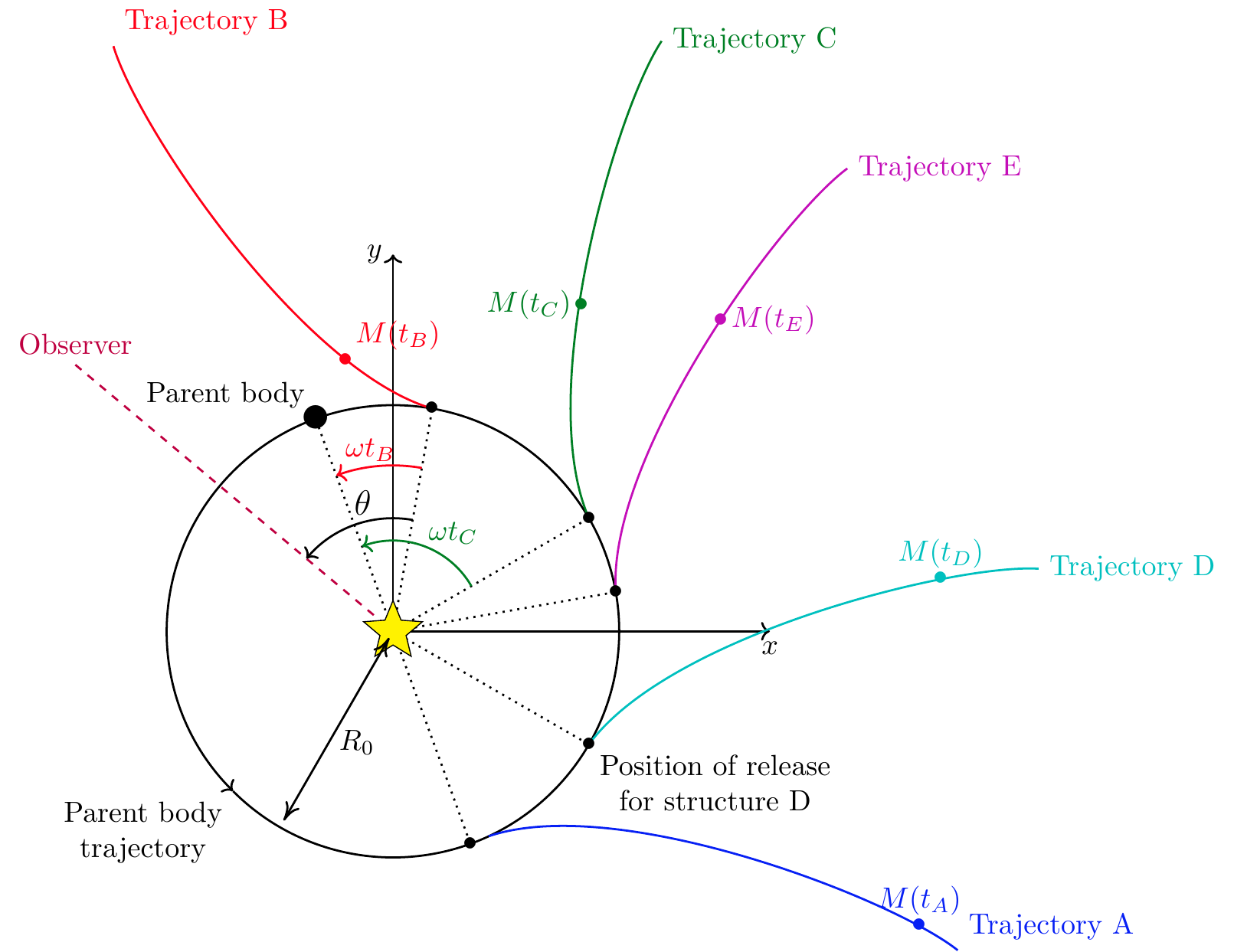}
}
}
\caption{\textbf{Left}: Sketch of the static case, seen from
  above. \textbf{Right}: Sketch of the orbiting case, seen from above.
  In this case, the structure B is used as the reference structure.
  Note that the features are not necessarily emitted in the alphabetical order.}
\label{fig:schSys}
\end{figure*}

\begin{table*}[tp!]
\caption{Apparent separations from the star (in arcseconds) 
  of the maximum elevation from midplane of the five arch-like structures,
  named A to E, observed in the AU~Mic debris disk in 2010, 2011 and 2014. 
  In 2010, the A structure is angularly too close to the star for being detectable.
  For consistency among all the features, 
  the uncertainties on the positions of the features A and B in 2011 
  have been re-estimated to twice their values, 
  as published in \cite{Boccaletti2015}.
  The next four rows give the positions of brightness enhancements 
  (and not the maximum elevation positions) identified in 2004 
  and {\it a posteriori} associated to the D and E structures. 
  The last row shows the mean positions and uncertainties on the structures seen in 2004.}
\label{tabPos}
\begin{tabular}{l l l l l l l}
\hline
Date	& A		& B		& C		& D		& E		& Reference\\
\hline
2014.69	& 1.017 $\pm$ 0.025 & 1.714 $\pm$ 0.037 & 2.961 $\pm$ 0.073 & 4.096 $\pm$ 0.049 & 5.508 $\pm$ 0.074 & \cite{Boccaletti2015}\\
2011.63 & 0.750 $\pm$ 0.025	& 1.384 $\pm$ 0.025	& 2.554 $\pm$ 0.025	& 3.491 $\pm$ 0.025 & 4.912 $\pm$ 0.208 & \cite{Boccaletti2015}\\
2010.69	& - 				& 1.259 $\pm$ 0.037	& 2.459 $\pm$ 0.049	& 3.369 $\pm$ 0.061 & 4.658 $\pm$ 0.245 & \cite{Boccaletti2015}\\ \hline \hline
2004.75	& -	& -	& -	& 2.52	& 3.22  & \cite{Fitzgerald2007}\\
2004.58	& -	& -	& -	& 2.52	& 3.12		& \cite{Liu2004}\\
2004.51	& -	& -	& -	& 2.21	& 3.22		& \cite{Metchev2005}\\
2004.34	& -	& -	& -	& 2.62	& 3.32		& \cite{Krist2005}\\ \hline 
2004.545 $\pm$ 0.147	& -	& -	& -	& 2.468 $\pm$ 0.154 & 3.220 $\pm$ 0.071 \\
\hline
\end{tabular}
\end{table*}

\subsection{Parameters and numerical approach}
\label{subsecParamSimu}

We adopt the stellar parameters listed in table~\ref{tabCaracAUMic}.
The stellar mass is not precisely determined, and we will take
$M_{\star}$= 0.4~M$_{\odot}$, consistent with \cite{Schuppler2015} and
the previous literature.  The impact of the assumed stellar mass on
the results will be discussed in Sec.~\ref{sec:mass}.  
 For the wind speed, we adopt the value in the literature of 450~km/s,
assumed to be constant with the distance from the star \citep[see][]{Strubbe2006,Schuppler2015}. 

Once these values are set, the particles' trajectories are fully
determined by two parameters: the radius \R\ at which the grains are
released, and the pressure to gravitational force ratio, \be.  
 To keep the problem simple,  we
suppose that all particles are submitted to the same pressure force,
meaning that we consider only one particle size and a time-averaged
value.  
 The case of a range of \be\ values is discussed in Sec.~\ref{subsecDistrib}. 
In our model, we assume that the dust release process takes
place inside of the planetesimal belt that is found to be located at
35--40~au.  We consider 40 values of \R\ ranging from 3 to 42~au, with
a linear step of 1~au.  \be\ is dependent upon the stellar activity.
AU~Mic is supposed to be on active state 10\% of the time, with
several eruptions per day.  \cite{Augereau2006} found values of
\be\ ranging from 0.4 in quiet state to 40 in flare state, with a
temporal average value of typically 4 to 5 (see their Fig.~11).  In
our case, we consider 40 values of \be\ ranging from 0.3 to 35 with a
geometric progression by step of $\times 1.13$, and thus including
bound orbits. 
 It has been analytically demonstrated that considering a time-averaged value of \be\ 
does not change the dust dynamics \citep[see Appendix C of][]{Augereau2006}, 
and we have numerically checked this behaviour. 

For numerical purpose, we work on a grid of (\R, \be) values and
optimize the values of the  other  parameters to minimize a $\chi^2$.
The trajectories are initially calculated for each  pair  (\R, \be) on
the grid.  A 4$\uma{th}$ order Runge-Kutta integrator, with a fixed,
default time-step equal to one hundredth of a year is used. The time
resolution on the parent body orbit will nevertheless be reduced to
0.1~year for numerical purpose in the case of an orbiting parent
body. The calculation of a trajectory stops after two revolutions for
the particle or for the parent body (if the particle has an unbound
trajectory), or earlier if the dust particle goes further than 200~au
from the star.  Then the computed trajectories are rotated with
respect to the observer to account for projection effects.  
Another parameter $\theta$ is thus introduced, corresponding to the angle
between the release point of the particle and the line of sight.  

Two models are used in the following: a model that assumes that the
parent body is static, and another where the parent body is rotating.
In both cases, the parent body intermittently emits dust particles.
In the static parent body model, the source of dust is static with
respect to the observer as illustated in the left panel of
Figure~\ref{fig:schSys}. The particles are all emitted with the same
angle $\theta$ with respect to the observer, follow the same
trajectory and differ only by their release dates. 
In the other model, the parent body moves on its orbit, assumed circular, between each
dust release event. 
Thus the angle of observation $\theta$ is linked
to the release dates as shown in the right panel of
Figure~\ref{fig:schSys}.  
Two structures emitted with a time difference $\delta t$ 
will be seen at an angle of $\omega \, \delta t$ from each other 
where $\omega$ is the parent body angular velocity.  
We set apart a structure that we call \emph{reference structure}. 
The angle of observation $\theta$ is defined with
respect to this reference structure and all angles for the other
structures are then deduced from the emission date.

In summary, the two models have a total 8 independent parameters: \R,
\be, $\theta$ and five dust release dates (one for each structure).
For each fixed (\R, \be)  pair,  the code finds the position of the
parent body that best matches the observations documented in
Table~\ref{tabPos} by adjusting the angle $\theta$ and the dust
release dates.  This is done by minimizing a $\chi^2$ value that takes
into account the uncertainties on the positions, and also on the
observing date in the specific case of the 2004 observations.


\section{Results}
\label{secResults}

We use the model described in the previous section to reproduce the
apparent positions of the five structures observed at three epochs:
2010, 2011 and 2014, see Table~\ref{tabPos}. We do not consider at
this stage the 2004 observations because the positions of the
structures have not been derived using the same approach as for the
other epochs. The consistency of our findings with the 2004 data is
discussed in Sec.~\ref{secDiscuss}. We first consider the simple case
of a static parent body (Sec.~\ref{subsecStatic}).  Then, we assume
the parent body is revolving on a circular orbit around AU~Mic
(Sec.~\ref{subsecOrbit}).

\subsection{Static parent body}
\label{subsecStatic}

\subsubsection{Nominal case}
\label{subsecNomSta}

\begin{figure*}[htp!]
\hbox to \textwidth{
	\parbox{0.5\textwidth}{
		\includegraphics[width=0.5\textwidth, height=!,trim=0 .6cm 0 .6cm]{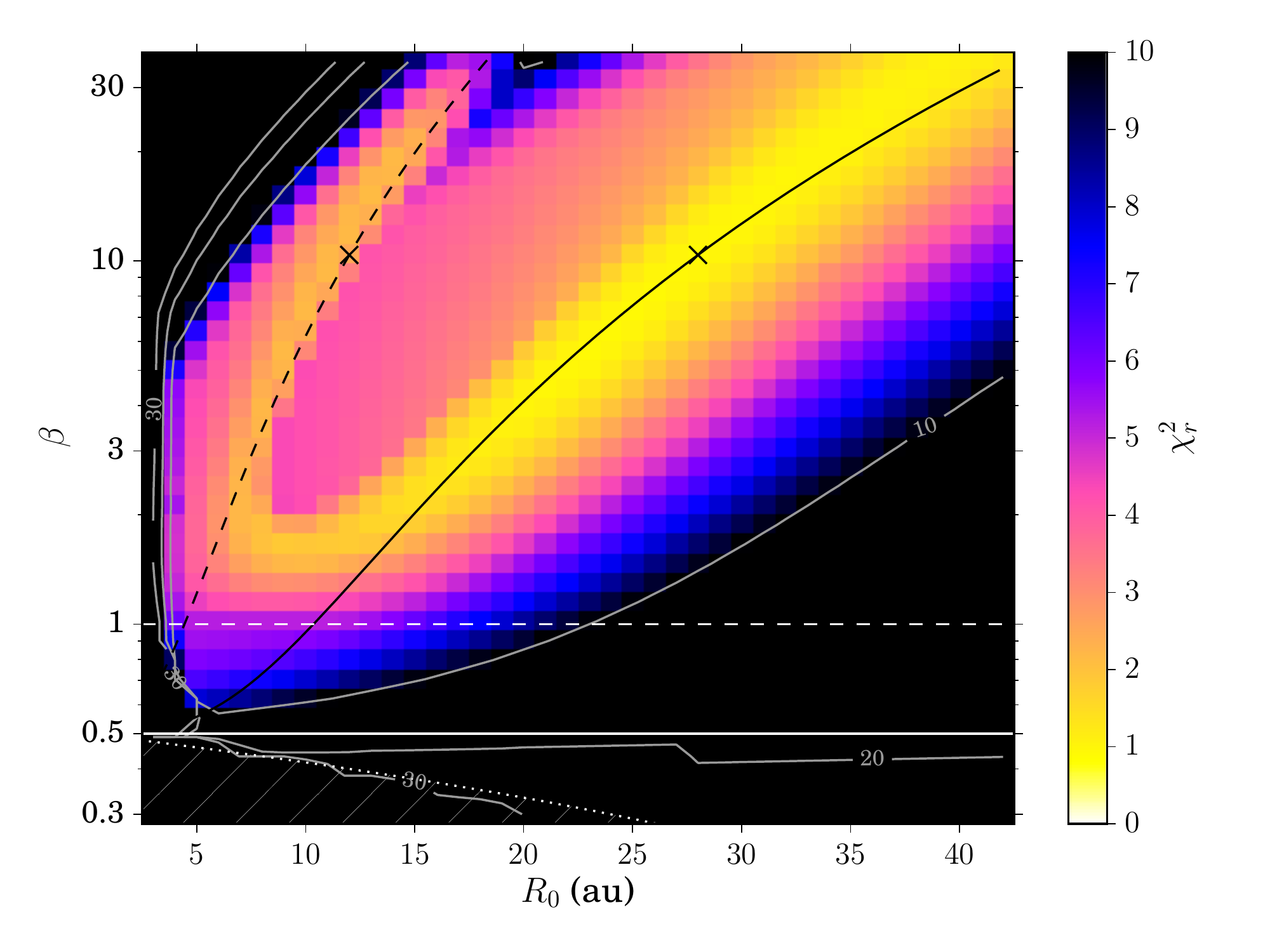}
		\subcaption{Mean \XX\ map of the fit to five structures}
		}
	\parbox{0.5\textwidth}{
		\includegraphics[width=0.5\textwidth, height=!,trim=0 .6cm 0 .6cm]{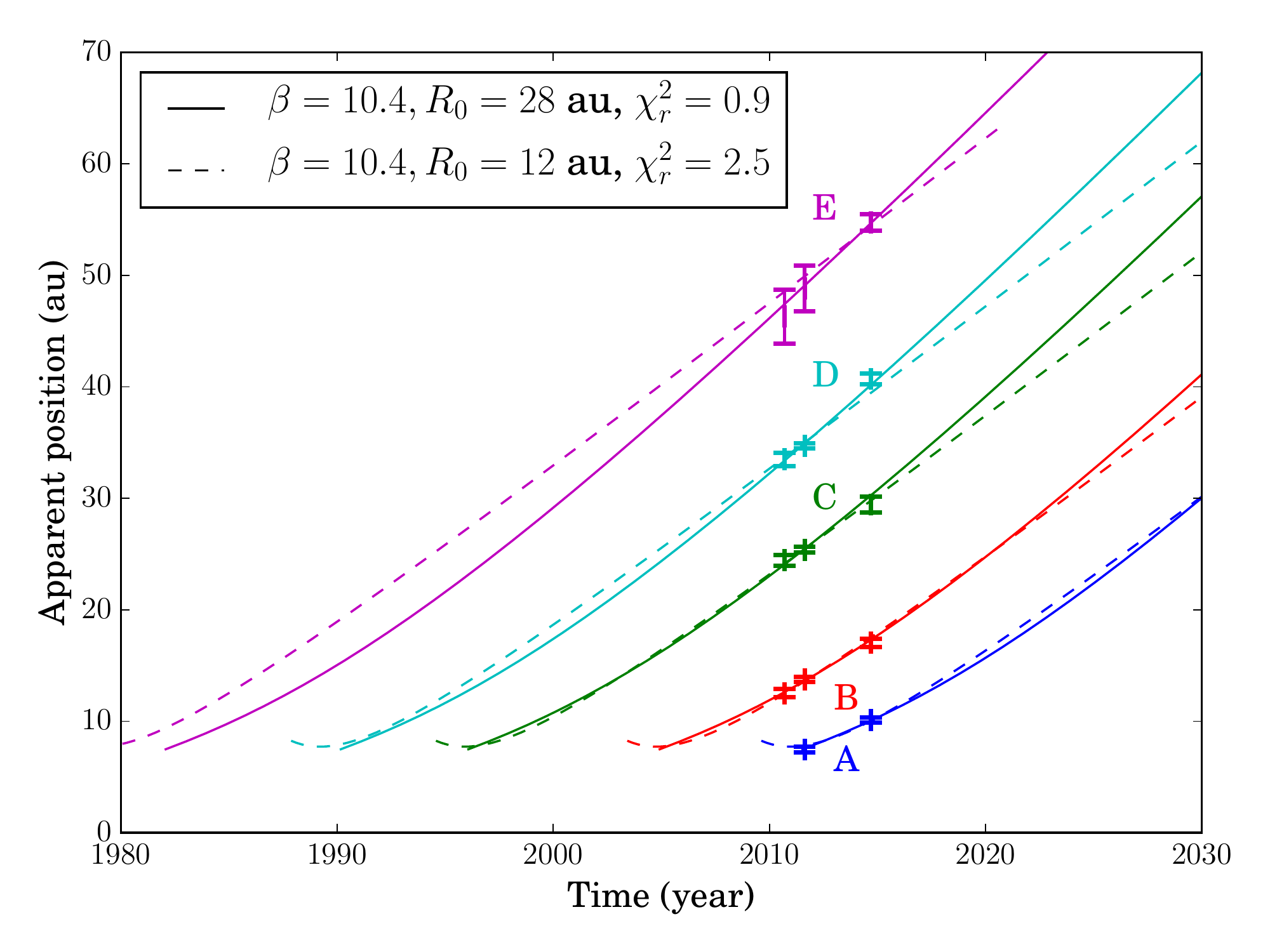}
		\subcaption{Best fits to the apparent positions as a function of observing date}
		}
	}
\hbox to \textwidth{
	\parbox{0.5\textwidth}{
		\includegraphics[width=0.5\textwidth, height=!,trim=0 0.6cm 0 0.2cm]{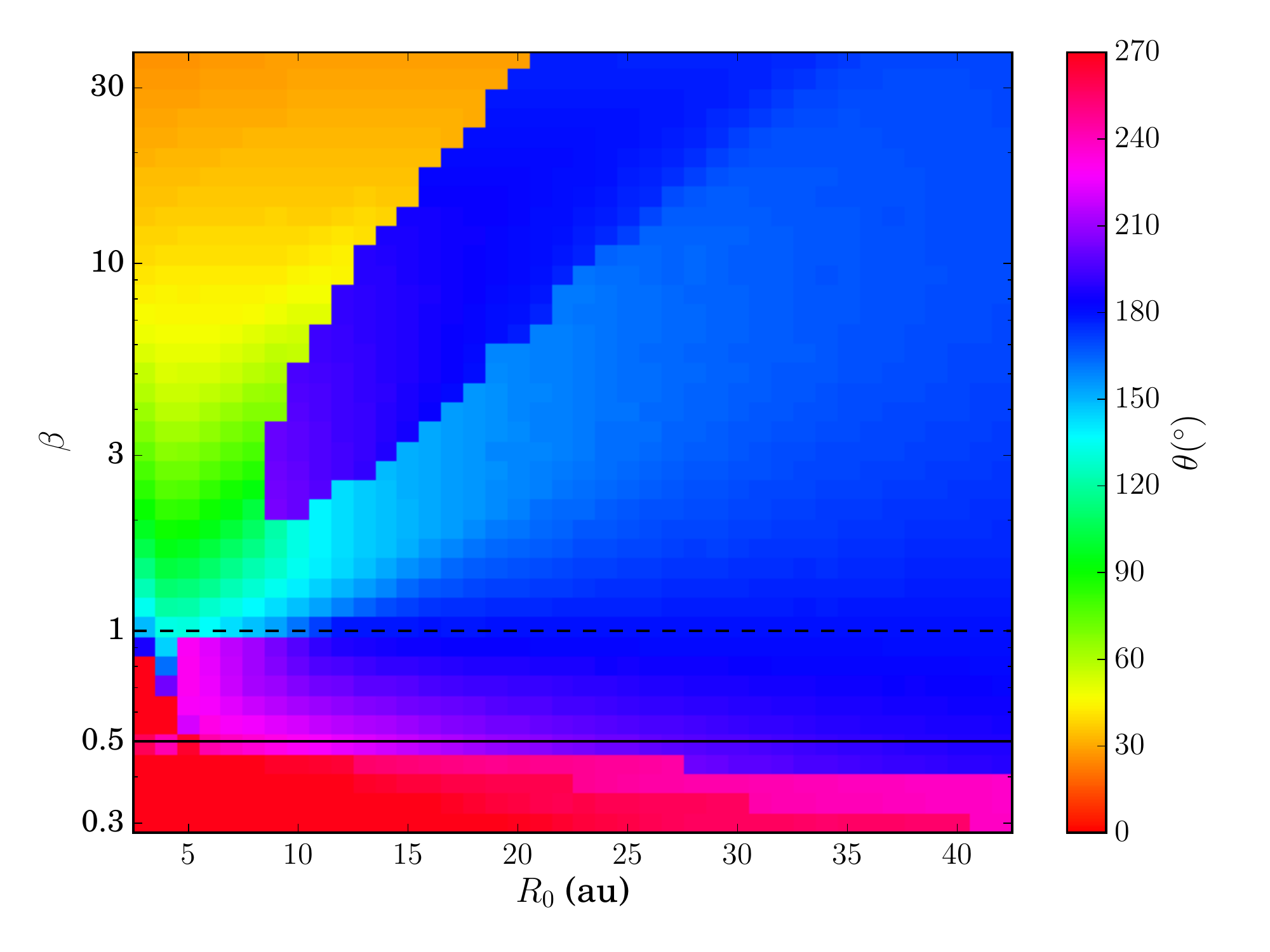}
		\subcaption{Map of the angle of emission with respect to the observer.}
		}
	\parbox{0.5\textwidth}{
		\includegraphics[width=0.5\textwidth, height=!,trim=0 0.6cm 0 0.2cm]{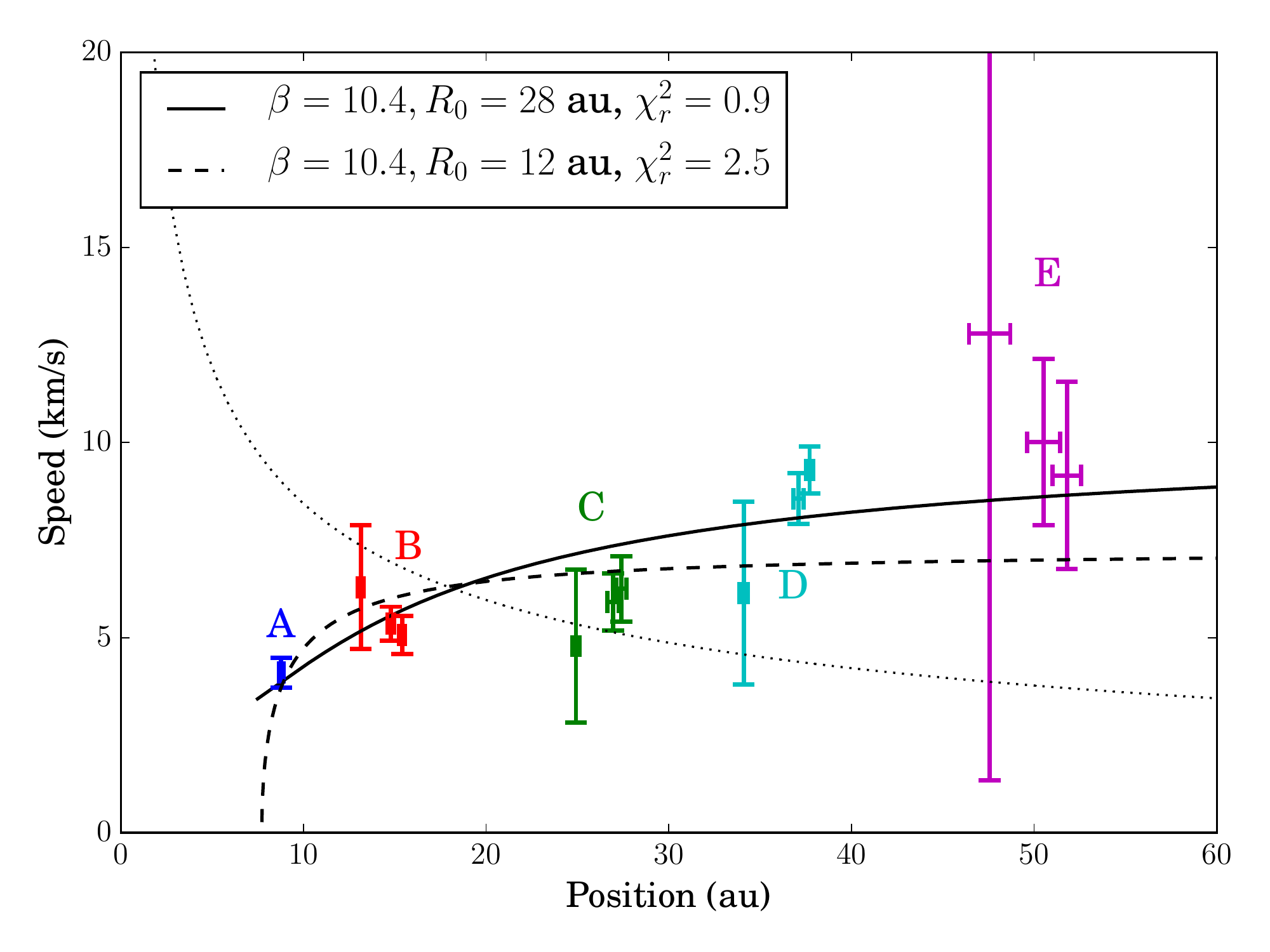}
		\subcaption{Best fits to the velocities as a function of apparent position}
		}
	}
\caption{\it Static parent body. \rm Modeling results in the case of a
  static parent body.  \textbf{Upper left}: Map of the reduced
  $\chi^2$ (\XX) obtained by fitting  the position of  the five structures.
  The solid and dashed black lines show the expected power law trends
  (Eq.~\ref{eqnAnaBR0}) along the two families of solutions, each
  scaled to go through the best fits identified by the black crosses.
  The branch of largest \R\ values corresponds to trajectories going
  away from the observer, while the other branch identifies the
  solutions pointing to the observer (see also
  Fig.~\ref{figAboStaABCDE}).
  The area under the dotted white line are solutions excluded by Eq.~\ref{eqnNoSol}.  
  \textbf{Bottom left}: Map of the angle of emission, taking with respect to the observer.
  \textbf{Upper Right}: Projected positions as a function of time. 
   The solid lines correspond to trajectories going away from the observer, while the dashed lines are trajectories pointing toward the observer
  (see also deprojected trajectories seen from above in Fig.~\ref{figAboStaABCDE}). 
  \textbf{Bottom right}: Same as Fig.~\ref{figVitObs}, overlaid with the best fit solutions (black crosses in panel (a)). 
  }
\label{figSpeAdjABCDE}
\end{figure*}

\begin{figure}[htp!]
\centering
\includegraphics[width=0.5\textwidth, trim=0 0.6cm 0 0.5cm,clip]{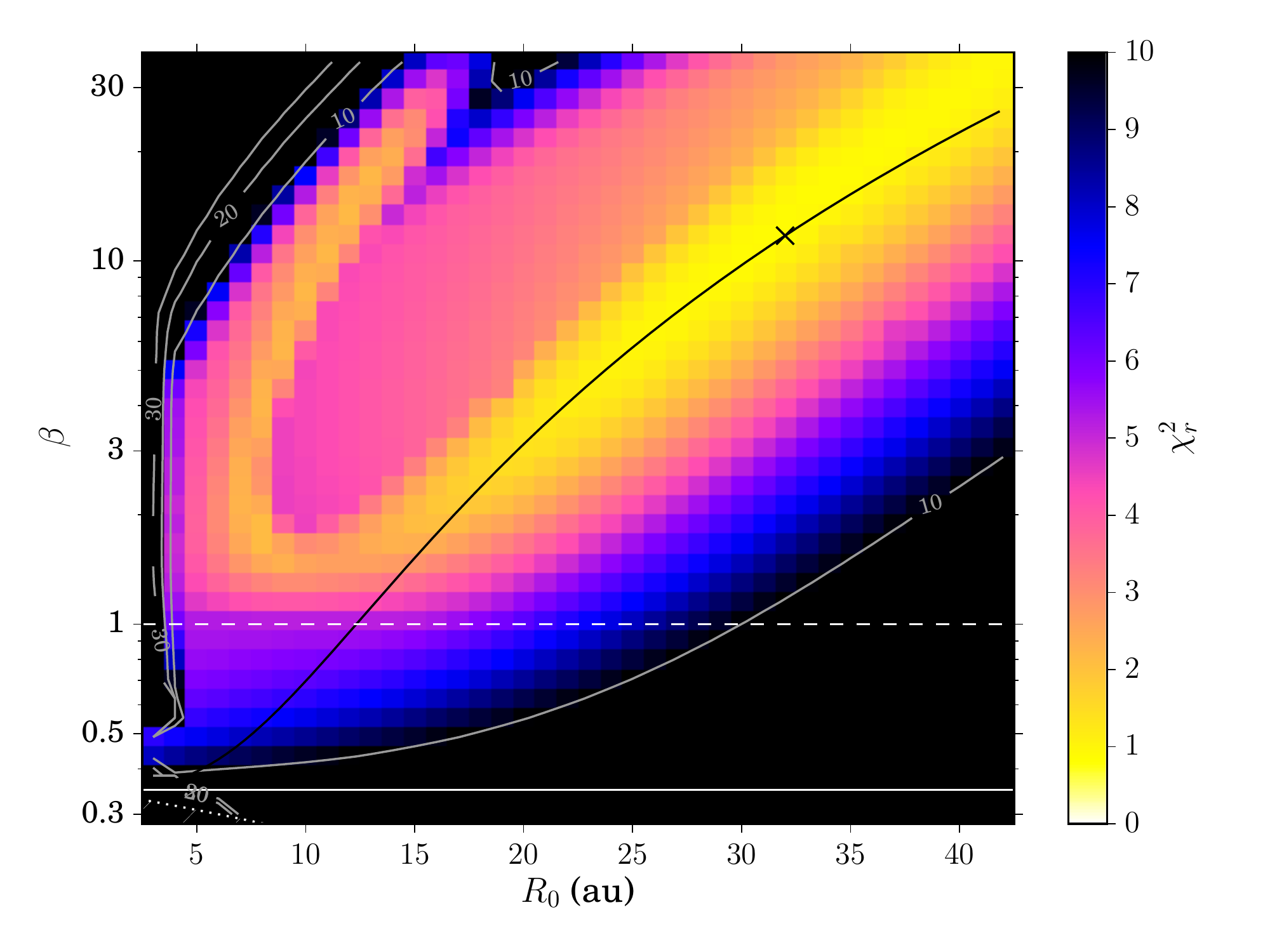} 
\caption{\it Static parent body. \rm Mean \XX\ map for a parent body with an eccentricity
    of 0.3.}
\label{figMapEcc}
\end{figure}

The simplest case to consider is that of a static parent body with
active periods during which it releases dust particles.  The reduced
\XX\ map of the fit to the apparent positions of the five structures
over time is displayed in Figure~\ref{figSpeAdjABCDE}a.  It shows two
branches of solutions, that both follow the expected trend, namely
\be\ raising as $R_0^3$ (Eq.~\ref{eqnAnaBR0}, solid and dashed black
lines in Fig.~\ref{figSpeAdjABCDE}a).  As can be seen in
Figure~\ref{figSpeAdjABCDE}c, the branch of solutions with the
smallest \R\ values corresponds to particles expelled out from the
AU~Mic system toward the observer ($0\degr \leq \theta \leq 90\degr$),
while the branch with the largest \R\ values, that also contains the
smallest \XX\ values, corresponds to grains moving away from
the observer ($90\degr \leq \theta \leq 180\degr$).  This is
illustrated in Figure~\ref{figAboStaABCDE}.  The best fit is obtained
for the \be~$\simeq$~10.4 bin of the grid, corresponding to particles
on unbound, abnormal parabolic trajectories as anticipated in
Sec.~\ref{subsecBehaviour}.

The likeliest values of \R\ and \be\ are derived using a statistical
inference method, by first transforming the map of unreduced $\chi^2$
into a probability map assuming a Gaussian likelihood function
($\propto \exp(-\chi^2/2$)), and then by obtaining marginalised
probability distributions for the parameters by projection onto each
of the dimensions of the parameter space (see for example Figs.~\ref{figDistribBeta} and \ref{figDistribRay}
in the orbiting case).  
 This gives $\beta = 10.5^{+21.6}_{-4.5}$, $R_0 = 28.4^{+7.9}_{-6.8}$~au 
and $\theta = 165\pm 6\degr$ (1$\sigma$ uncertainties).   
The simulation closest to these values in the grid of models (right black cross in
Fig.~\ref{figSpeAdjABCDE}a, \XX = 0.9) is shown in
figures~\ref{figSpeAdjABCDE}b and \ref{figSpeAdjABCDE}d, and the
release dates of particles are documented in Table~\ref{tabEmi}.  It
shows a quasi-periodic behaviour of about 7~years, with the structures
the closest to the star in projection being the
youngest. Interestingly, we note that in this model, the dust forming
the A structure is released in mid-2011, which would be consistent with
the non-detection of that feature in the 2010 HST/STIS data.

\begin{figure*}[htp!]
\centering
\hbox to \textwidth
{
\parbox{0.5\textwidth}{
\includegraphics[width=0.5\textwidth, height=!,trim=0 .6cm 0 .6cm]{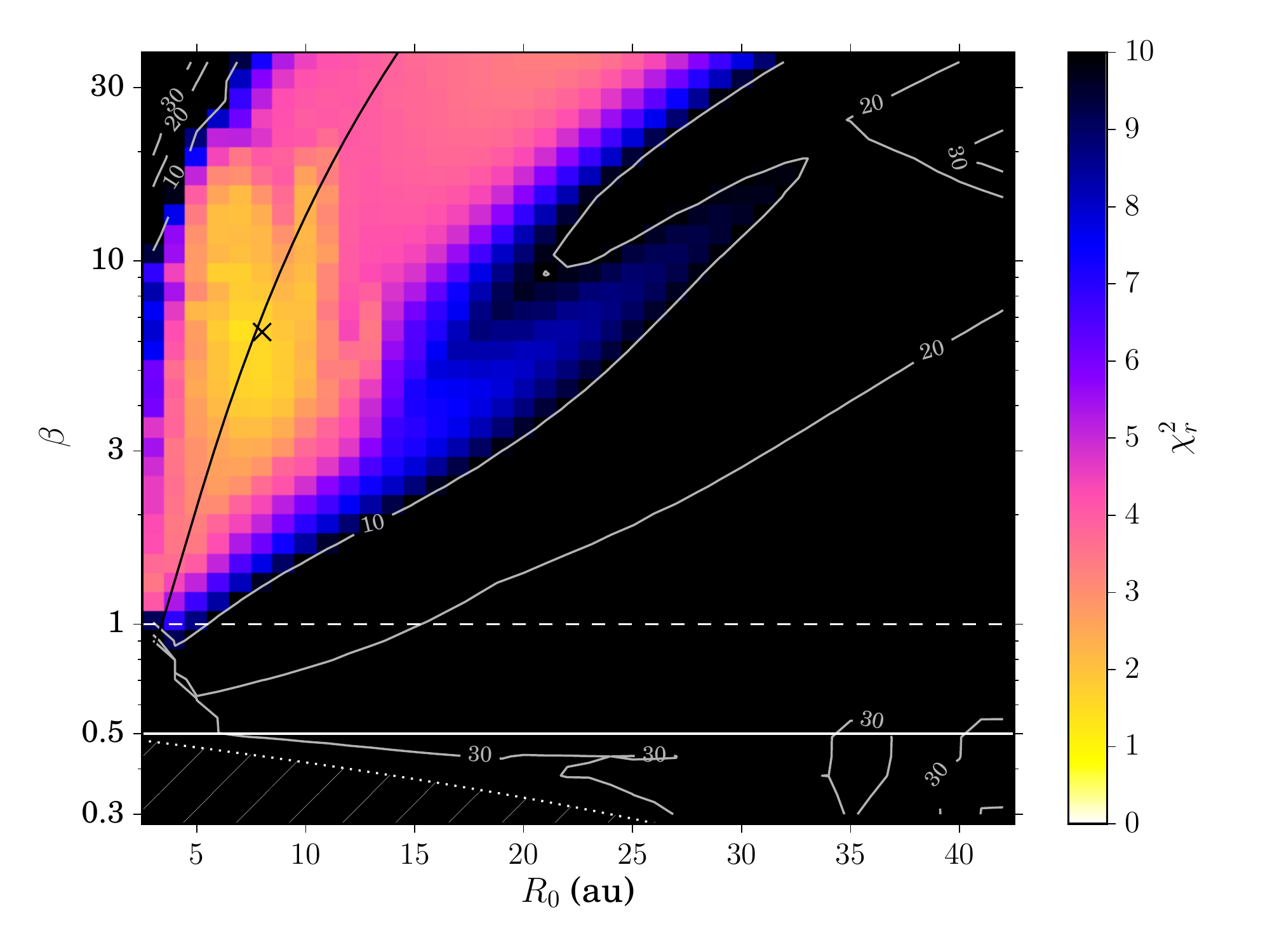}
\subcaption{Mean \XX\ of the fit to five structures}
}
\parbox{0.5\textwidth}{
\includegraphics[width=0.5\textwidth, height=!,trim=0 .6cm 0 .6cm]{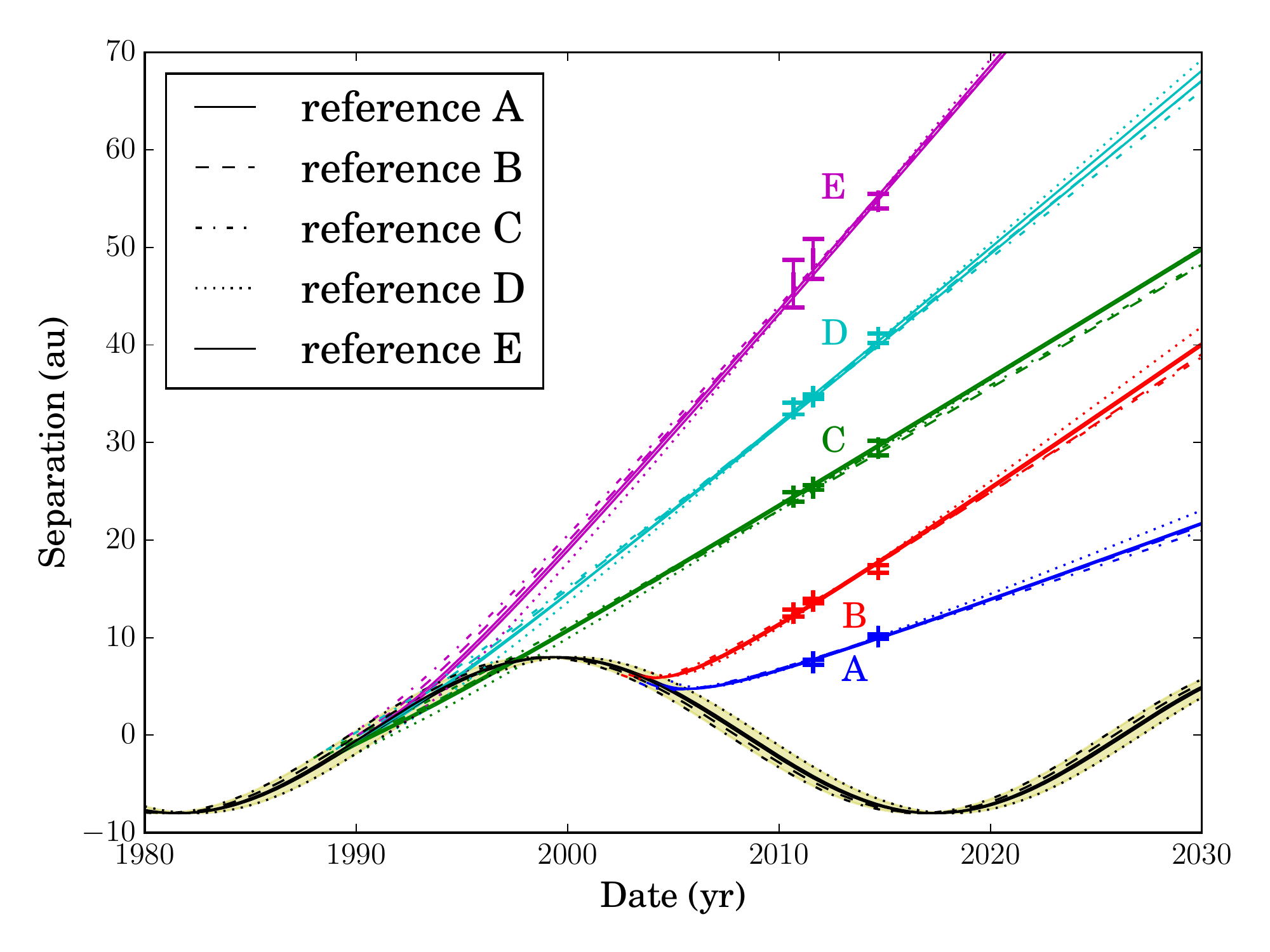}
\subcaption{Best fits to the apparent positions as a function of observing date}
}
}
\caption{\it Orbiting parent body. \rm Modeling results of the position adjustement of all the
  structures in the case of an orbiting parent body (nominal case, see Sec. \ref{subsecNomOrb}).
  \textbf{Left}:
  \XX\ map averaged over the five references (see
  text). \textbf{Right}: projected trajectories for the five
  structures assuming each of the structure is a reference for the
  fit. This shows that the fits are essentially independent of the
  assumed reference structure.
  The black lines are the apparent positions of the parent body.}
\label{figPosMoy}
\end{figure*}

\subsubsection{Eccentric orbits}
\label{subsecEcc}

At first glance, the case of a static parent body might appear a less
physical situation than the case of an orbiting parent body. It could
nevertheless correspond to a high density region of large velocity
dispersion in the aftermath of a giant collision. As shown by
\citet{Jackson2014} for example, the collision produces a swarm of
large objects, passing through the same position in space, that will
in turn become the parent bodies of the observed dust grains. This
could mimic a static parent body, but importantly, the grains may be
released from parent bodies on eccentric orbits. This will affect
their initial velocity. Therefore, we test the impact of the parent
body's eccentricity on the results by considering dust particles
released at the pericenter position of parent body's orbit. 

We arbitrarily consider parent bodies with an eccentricity of
$e=0.3$. The corresponding \XX\ map is displayed in
Figure~\ref{figMapEcc}b. The eccentricity lowers the limit between
bound and unbound trajectories in terms of \be.  The total energy per
mass unit becomes $e_m = \frac{GM_{\star}}{2 R_0} (2\beta -1 +e)$.  In
our case, the bound trajectories correspond to $\beta < 0.35$.  The
limit between normal and abnormal parabolic trajectories stay the
same, \be\ = 1.  The power law of equation~\ref{eqnAnaBR0} is also
modified, leading to $(2\beta -1 +e) \propto R_0^3$. Introducing an
eccentricity globally improves the fits (lower \XX) at small
\be\ values, but does not change significantly the best solutions.
 The likeliest values of \be\ and \R\ are respectively
$12.1^{+17.7}_{-4.5}$ and $31.8^{+7.9}_{-4.8}$~au.


\begin{table*}
\caption{Release dates (in years) of the structures A to E for the best fit for all the models considered.
  The last row shows the average over the five reference structures.
  Uncertainties are derived from the dispersion of the results. }
\label{tabEmi}
\begin{center}
\begin{tabular}{| c| c|  c|  c|  c|  c |}
\hline
	& A		& B		& C		& D		& E \\
\hline
\multicolumn{6}{| c |}{~} \\
\multicolumn{6}{| c |}{Static: $\beta = 10.4, R_0 = 28$~au} \\
\hline
\input{tabLatexStatic.tex}

\multicolumn{6}{| c |}{~} \\
\multicolumn{6}{| c |}{Orbiting free: $\beta = 6.4, R_0 =8$~au} \\
\hline
\input{tabLatexFree.tex}

\multicolumn{6}{| c |}{~} \\
\multicolumn{6}{| c |}{Orbiting frontward: $\beta = 24.4, R_0 = 17$~au} \\
\hline
\input{tabLatexFront.tex}

\multicolumn{6}{| c |}{~} \\
\multicolumn{6}{| c |}{Orbiting backward: $\beta = 5.6, R_0 = 8$~au} \\
\hline
\input{tabLatexBack.tex}

\end{tabular}
\end{center}
\end{table*}



\subsection{Orbiting parent body}
\label{subsecOrbit}

\subsubsection{Nominal case}
\label{subsecNomOrb}

\begin{figure*}[htp!]
\centering
\hbox to \textwidth
{
\parbox{0.33\textwidth}{
\includegraphics[width=0.33\textwidth, trim=4.8cm 0.5cm 5.9cm 0.4cm, clip]{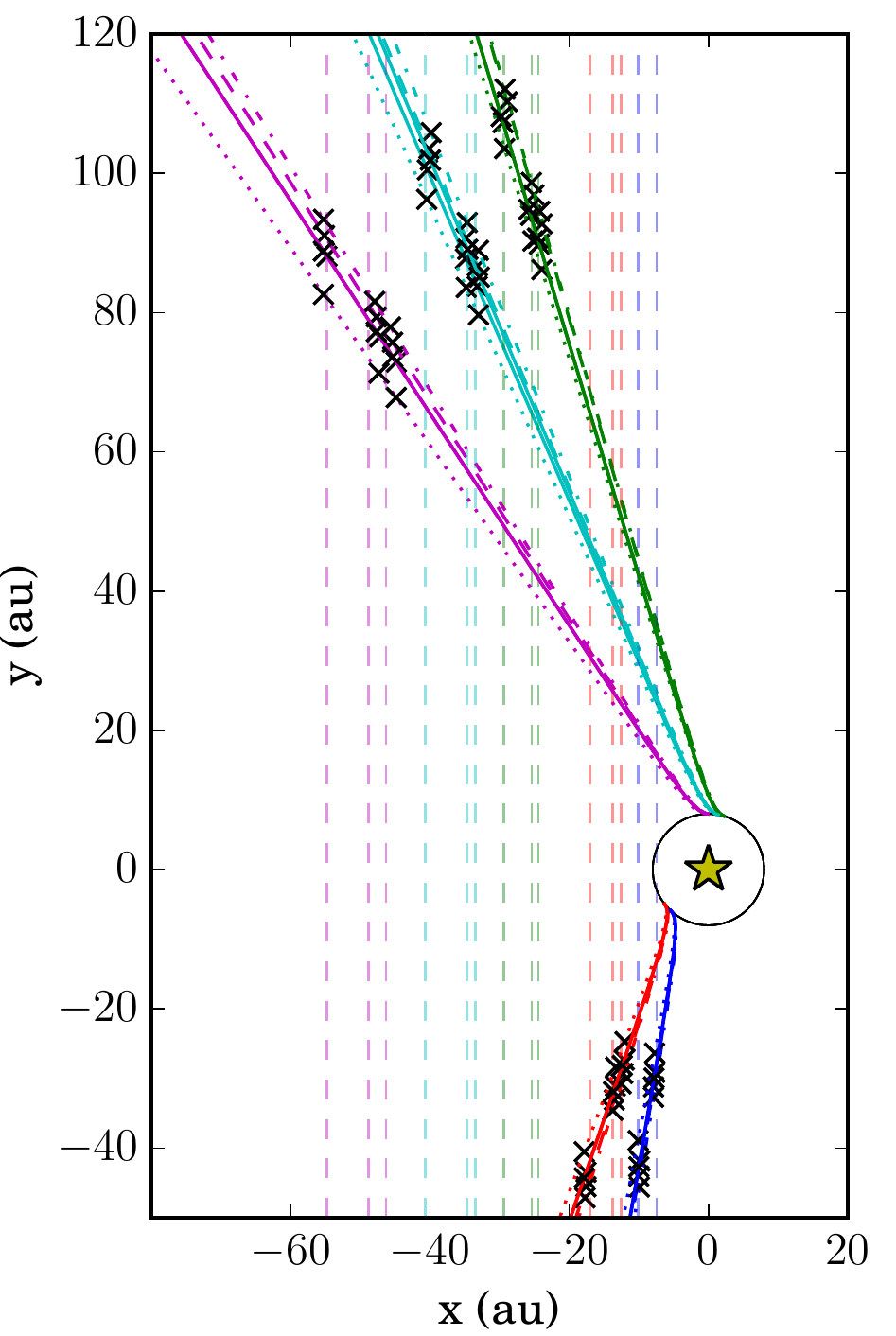}
\subcaption{Model without constraint}
}
\parbox{0.33\textwidth}{
\includegraphics[width=0.33\textwidth, trim=4.4cm 0.5cm 6.1cm 0.2cm, clip]{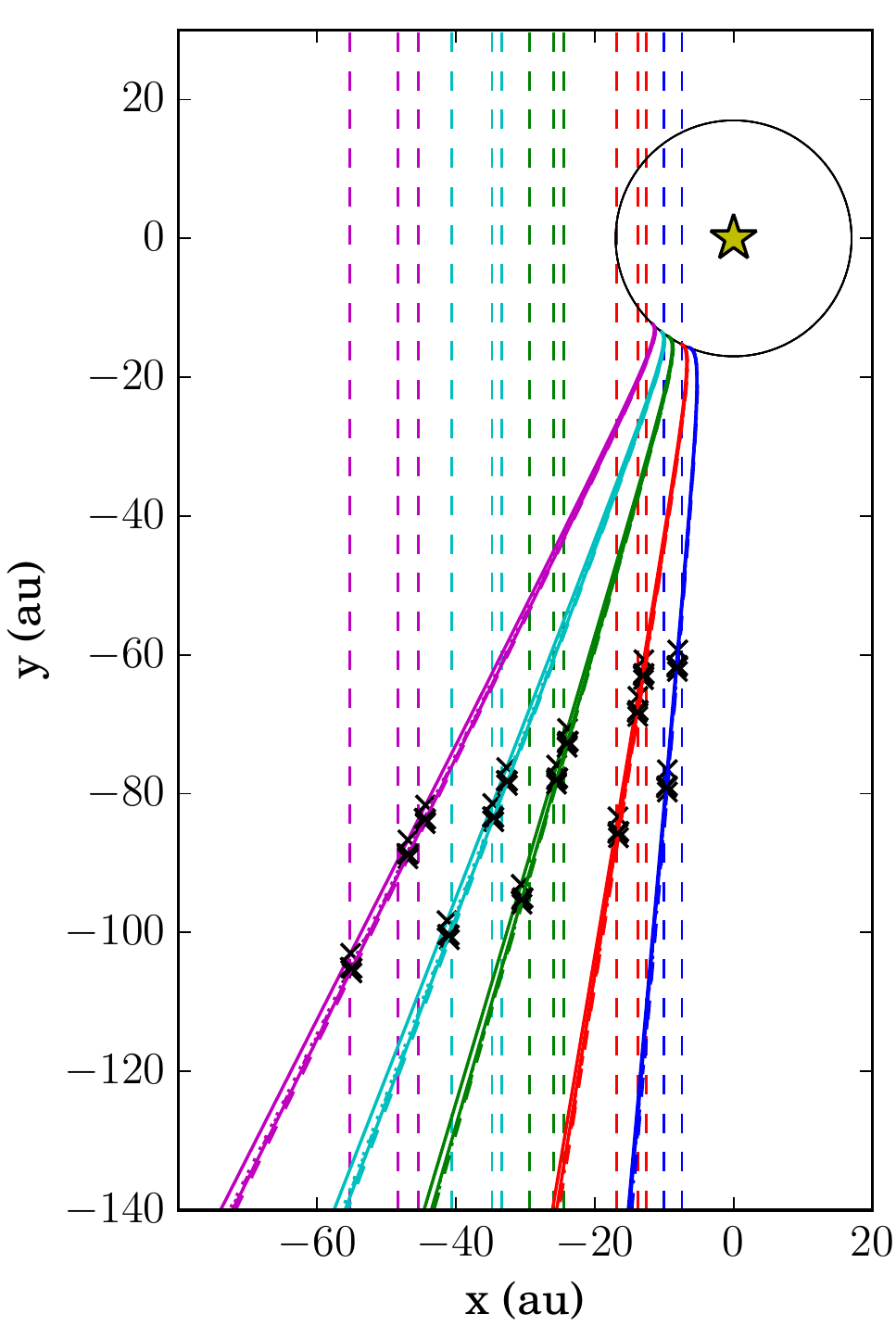}
\subcaption{Particles going toward the observer}
}
\parbox{0.33\textwidth}{
\includegraphics[width=0.33\textwidth, trim=4.8cm 0.5cm 5.9cm 0.4cm, clip]{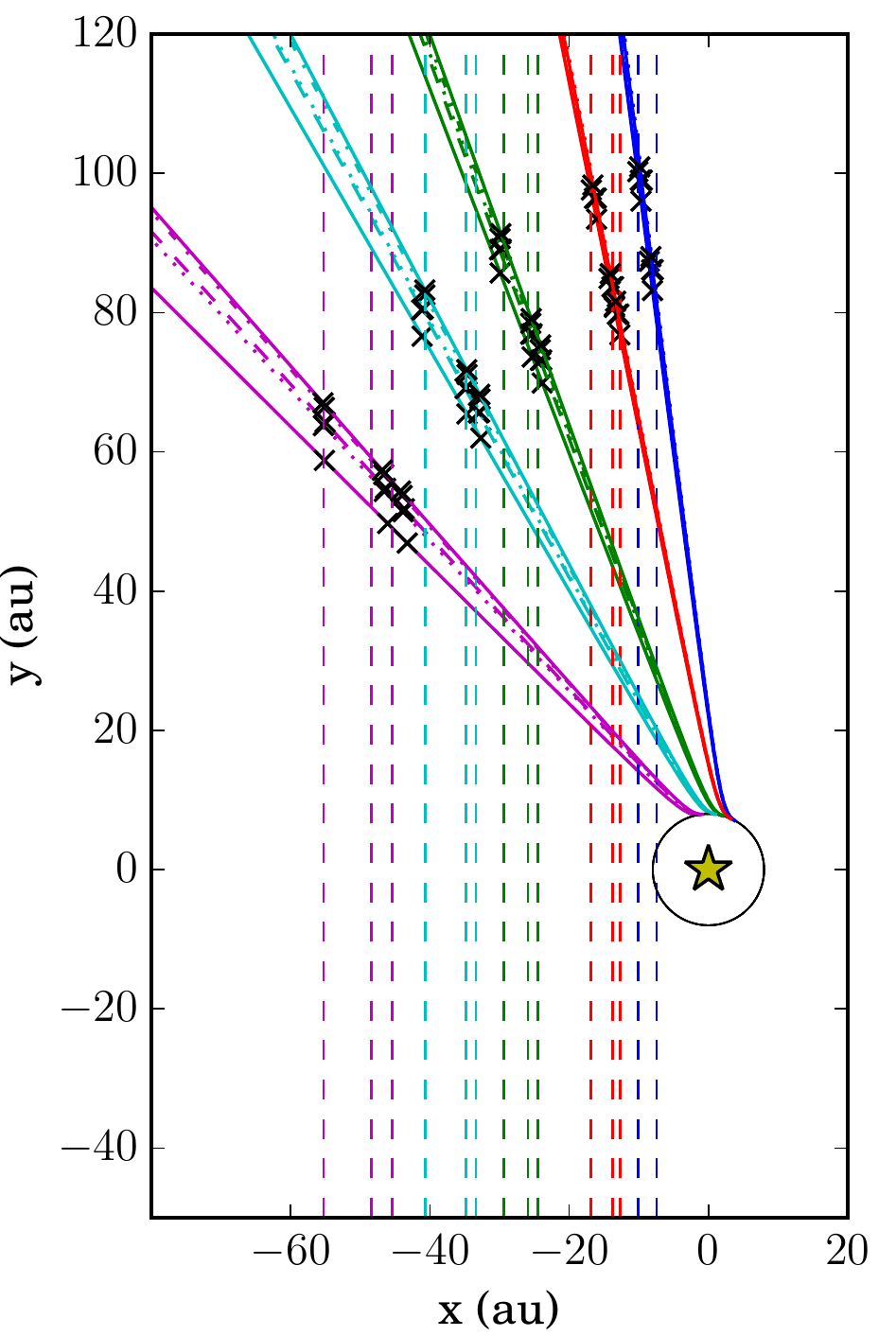}
\subcaption{Particles going away from the observer}
}
}
\caption{\it Orbiting parent body. \rm
Trajectories of the particles seen from above for the five
  structures in the case of a rotating parent body (see
  Secs.~\ref{subsecNomOrb} and \ref{sec:grouped} for details). For each
  structure, five similar trajectories, sometimes superimposed, are
  displayed corresponding to best fits obtained when the reference
  structure is varied in the model. The color-coding is the same as in
  previous figures. The solid black circle is the trajectory of the
  parent body. The crosses correspond to the observing dates.}
\label{figPosMoyAbo}
\end{figure*}

We now consider the case of a parent body on a circular orbit. We
assume an anti-clockwise orbit when the system is seen from above as
illustrated in Fig.~\ref{figPosMoyAbo} for example, but it was
numerically checked that considering a clockwise orbit yields similar
results, as expected (the $x$-axis is an axis of symmetry for the
problem). The five structures are supposed to correspond to activity
periods, when dust is released, occuring at different positions of the
parent body on its orbit (Fig.~\ref{fig:schSys}, right
panel). Therefore, each structure has its own trajectory although
these are all self-similar in shape because they share the same
\R\ and \be\ values. For each (\R, \be)  pair,  we adjust the observed
positions of the structures as a function of time, alternately
considering each of the five structures as a reference structure in
the fitting process (see Sec.~\ref{subsecParamSimu} for details). This
yields five fits to the data for any (\R, \be)  pair,  that appeared
to be consistent with each other, although with slight differences,
and the results were averaged to derive a single \XX\ map.

The results for an orbiting parent body are shown in Figure~\ref{figPosMoy}. 
The \XX\ map evidences a region of best fits with \be\ values similar to those found in the case of a static parent body, 
but for a dust release source much closer-in. 
 The likeliest values derived using the statistical inference method described in Sec~\ref{subsecStatic} 
are $\beta = 6.3^{+3.0}_{-2.4}$ and $R_0 =7.7^{+1.0}_{-1.5}$~au. 
The closest solution in our grid of models is represented by the black cross on left panel of Fig.~\ref{figPosMoy},
namely $\beta = 6.4$ and $R_0 = 8$~au ($\chi^2_r = 1.7$).  
The corresponding projected trajectories for the five structures are displayed in the right panel of Fig.~\ref{figPosMoy}, 
showing an excellent agreement with the observations, independent of the reference structure used. 
We note, however that the solutions are very close in terms of \XX , 
and that additional observations are necessary to constrain the trajectory better.

From these results, we can derive a dust release date for each of the
structures for the best fit model.  These are listed in
Table~\ref{tabEmi} (labelled "Orbiting free"), 
where the uncertainties combine the dispersion on
the best ten percent  pairs  for a given reference structure, and the
dispersion within the fits with the five different reference
structures.  
 In this model, the C structure appears first (in $\sim$1989), 
followed by D and E with an almost 1-year periodicity. 
These three trajectories point in a direction opposite to the
observer.  
Structures A and B, on the other hand, are released much
later, early 2000, about 10 to 15~years after structure E. Their
trajectories are furthermore oriented toward the observer. A face-on
view of the five trajectories is displayed in the left panel of
Figure~\ref{figPosMoyAbo}.

 Although the 1-year periodicity for the C to E structures could
provide some hints on the origin of the dust release process, the
specific behaviour of the A and B structures is calling for staying cautious
as about the interpretation of the model. 
This motivates us to test in
the following the case of grouped release events for all the
structures, in a time span shorter than a quarter of the parent body
orbital period.

\subsubsection{Grouped release events}
\label{sec:grouped}

We keep exploring the case of an orbiting parent body, but we now
force the structures to be emitted more closely in time than
previously. This is numerically achieved by limiting the accessible
range of dust release dates to a quarter of the parent body orbital
period. This leads to considering two situations: the case of five
trajectories all oriented toward the observer on one hand, and five
trajectories all moving away from the observer in the other hand.

It turns out that none of these scenarii yields better fits to the data based on a \XX\ criterium, 
as one could anticipate since these situations where numerically considered in the nominal case (previous section). 
The best fit to the positions of the structures in time with
particles forced to be emitted in the direction of the observer is
shown in the middle panel of Fig.~\ref{figPosMoyAbo}. 
 It corresponds
to $\beta = 24.7^{+10.6}_{-2.9}$, $R_0 = 17.3^{+4.7}_{-3.0}$~au, and
the corresponding dust release dates are reproduced in Table~\ref{tabEmi}. 
The fits with different reference structures are consistent with each other.
The values of \be\ and \R\ are significantly
larger than those obtained in the nominal orbiting case
(Sec.~\ref{subsecNomOrb}). The upper limit on \be\ is in fact reaching
the upper bound of the explored range in our simulations
(see Fig.~\ref{figMapGroup}), and we
checked that expending this range increases the best \be\ value, as
well as the corresponding \R\ value in accordance to
Eq.~\ref{eqnAnaBR0}. 
 The reduced \XX\ of about 3.6 is worse than in the nominal orbiting case, 
but it is interesting to note that a dust
release periodicity of about 1.5 years does appear in this model, with the
structures at the largest projected distances from the star being the
oldest (release dates between about 1994 and 2000 for the E to A
structures, respectively). 

The case of particules forced to be pulled away from the observer
yields quite different results. 
 The best fit is obtained for $\beta =
5.6^{+4.8}_{-3.6}$ and $R_0 = 8.1^{+2.0}_{-3.1}$~au (see Fig.~\ref{figPosMoyAbo}c), with a reduced
\XX\ value around 3.5.  
In this case, the
dispersion in the parameter values due to the use of different
reference structures is larger than before, as can be seen on the
right panel of Fig.~\ref{figPosMoyAbo}. Overall, the mean \be\ and
\R\ values are similar than in the nominal case
(Sec.~\ref{subsecNomOrb}). The dust release dates are documented in
Table~\ref{tabEmi}.
 It shows that the periodicity is a little
smaller than one year and the closest structures in projection are the
oldest in this model, with structure A appearing in $\sim$1990 and the
last structure (E) in $\sim$1994. 

In summary, 
even if the grouped emission solutions are not the best based on the \XX\ criterium, 
they present the conceptual advantage of a periodicity.


\section{Discussion}
\label{secDiscuss}

\begin{figure*}[!]
\centering
\hbox to \textwidth
{
\hspace{-0.3cm}
\parbox{0.5\textwidth}{
\includegraphics[width=0.49\textwidth, height=!,trim=0 0.5cm 0  0.5cm,clip]{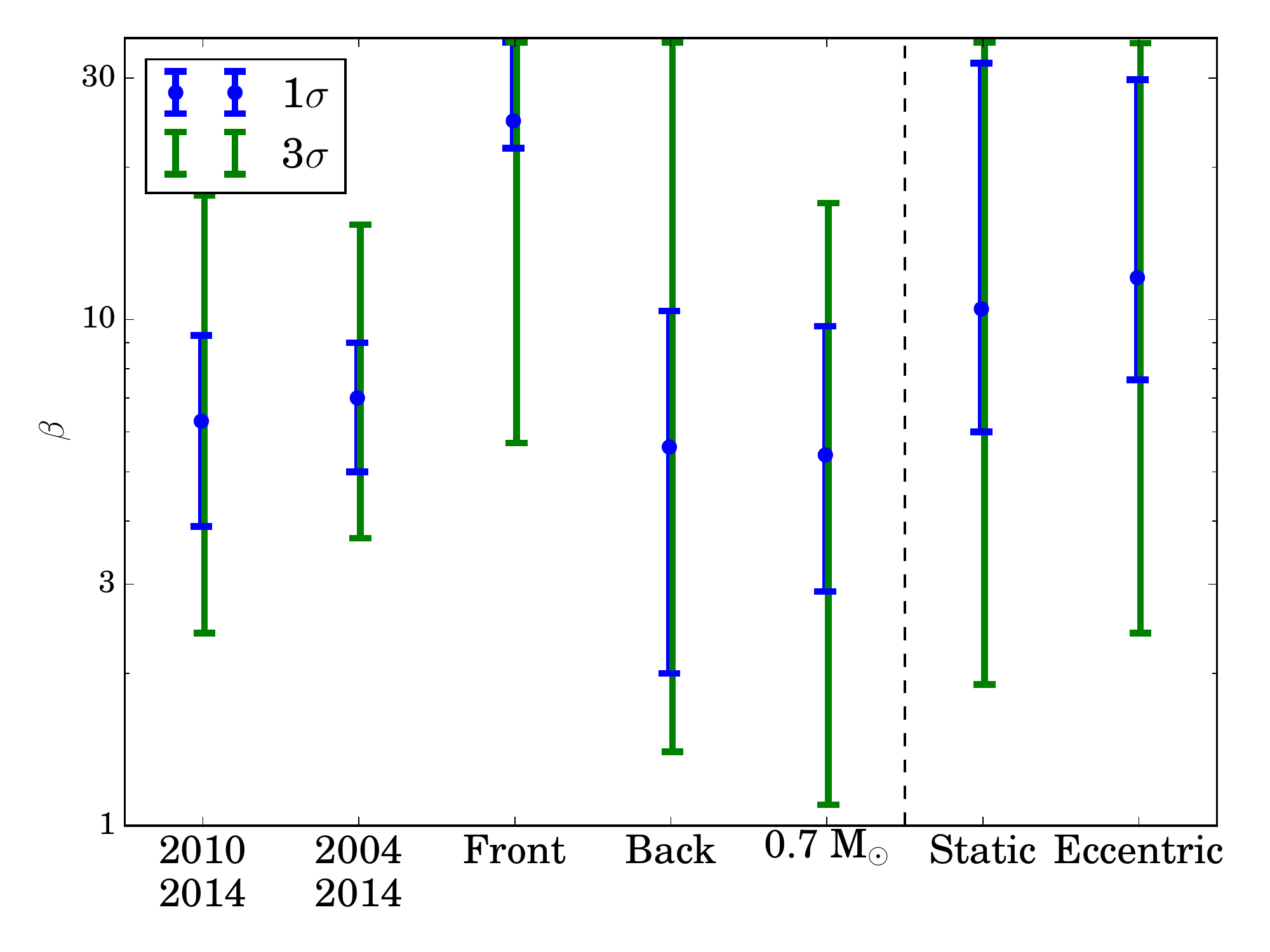}
}
\parbox{0.5\textwidth}{
\includegraphics[width=0.49\textwidth, height=!,trim=0 0.5cm 0 0.5cm,clip]{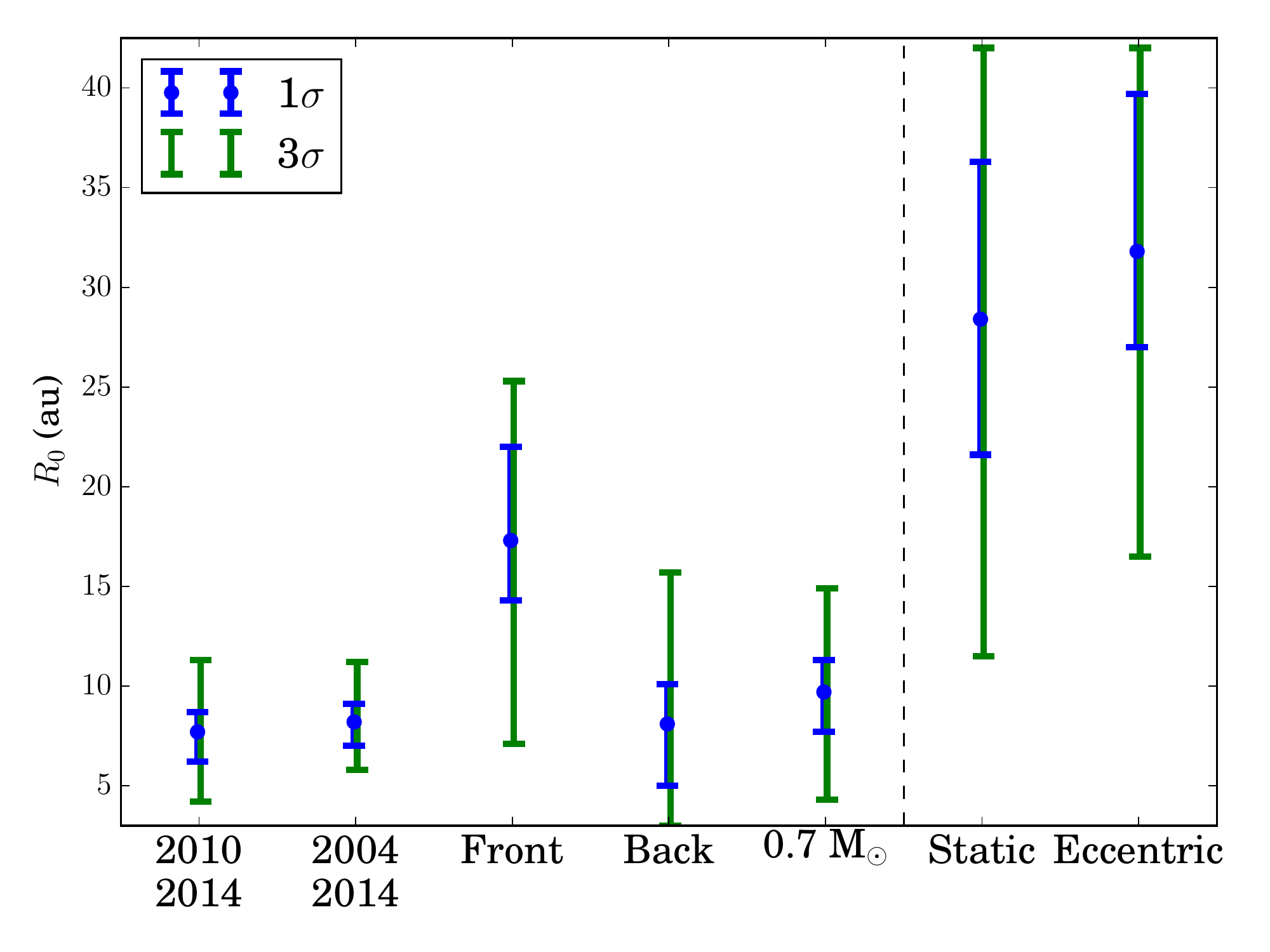}
}
}
\caption{\textbf{Left}: Likeliest values of \be\ depending on the
  model.  "2010 2014" is the nominal orbiting case
  (Sec.~\ref{subsecNomOrb}), "2004 2014" is the same model taking
  into account the 2004 observations (Sec.~\ref{sec:2004}), "0.7
  M$_{\odot}$" is the model with a stellar mass of 0.7 M$_{\odot}$
  (Sec.~\ref{sec:mass}), "Front" and "Back" are the grouped solutions
  (Sec.~\ref{sec:grouped}), "Static" is the model of a static parent
  body (Sec.~\ref{subsecNomSta}) and "Eccentric" is the model discussed
  in Sec~\ref{subsecEcc}. \textbf{Right}: Same for the dust
  release position \R.}
\label{figBetaRaySol}
\end{figure*}

\subsection{Comparison between models}

The simulations reproduce the general trend of increasing projected velocities 
of the structures with increasing distance to the star. 
This behaviour can be explained by an outward acceleration of the particles 
being pushed away by a stellar wind pressure force 
that significantly overcomes the gravitational force of the star. 
The static and nominal orbiting parent body models provide equally good fits to the data. 
Figure \ref{figBetaRaySol} provides a digest of the \be\ 
and \R\ values found in this study, 
along with the error bars at 1$\sigma$ and 3$\sigma$. 
Our model requires the stellar wind to be strong enough
to achieve \be\ values between typically 3 and 10.  
In the static case, the dust seems to originate 
from a location just inside the planetesimal belt, at 25--30~au from the star, 
while in the case of an orbiting parent body, 
the best fit model is obtained for a dust release distance to the star \R\ of about 8~au. 
The release dust events are less than 30 years old, 
dating back  to  the late 1980's for the oldest, 
while the most recent features would have been emitted 
in the mid-2000 at the latest in the case of an orbiting parent body, 
and as late as mid-2011 in the case of a static parent body. 
Some periodicity does appear in the simulations, 
but these depend on the model assumptions 
and current data are not sufficient to disentangle 
between the various scenarii considered in this study. 
For instance, the static parent body model shows a $\sim$7~year periodicity, 
while some 1 to 2~year periodicities are noticed 
when considering an orbiting parent body, 
with a possible 10-15 year inactivity period in the best fit model 
(Fig.~\ref{figPosMoyAbo}a).

In the case of an unconstrained orbiting parent body, 
the best fit model suggests that the C structure is older than the D structure 
that is itself older than the E structure (Tab.~\ref{tabEmi} and Fig.~\ref{figPosMoyAbo}a). 
The observations would naively suggest the opposite, 
namely that the closest structures are the youngest.
Indeed, the vertical amplitude of the arch-like structures seems to decrease 
with increasing apparent position \citep{Boccaletti2015}, 
suggesting for example a damping process when the structures move outwards. 
The observed increase of the radial extent of the arches would also support this conclusion, 
although projections effects could also explain this behaviour. 
In fact, independent of the scenarii displayed in Fig.~\ref{figPosMoyAbo}, 
the orientation of the trajectories with respect to the observer are such that, 
should the arches have the same shape, 
their apparent radial extent would increase 
with increasing projected distance to the star, as observed. 
This criterion does not allow to exclude one scenario, 
but ongoing follow up observations could constrain
the orientation of the structures with respect to the line of sight.

It is also worth mentioning that the case of grouped release events 
toward the observer (Fig.~\ref{figPosMoyAbo}b) does yield surprising results 
that must be taken with care. 
For this case, the best fits tend to be obtained for the largest possible \be\ values 
in our grid of models and extending the range of \be\ values does confirm this trend. 
 However, we note that the improvement in terms of \XX\ is limited, 
and that fixing for instance \be\ to about 6 would correspond to 
\R\ values close to 10\,au ($\chi_r^2=4.6$, see Fig.~\ref{figMapGroup}a), 
in better agreement with the other models. 
Therefore, in the following discussion, we will adopt $\beta = 6 \pm 1$ and $R_0 = 8 \pm 2$~au 
as representative values in the case of an orbiting parent body, 
independently of whether the release events are grouped or not.

\begin{figure*}[!]
\centering
\hbox to \textwidth
{
\hspace{-0.3cm}
\parbox{0.5\textwidth}{
\includegraphics[width=0.51\textwidth, height=!,trim=0 0.5cm 0  0.5cm,clip]{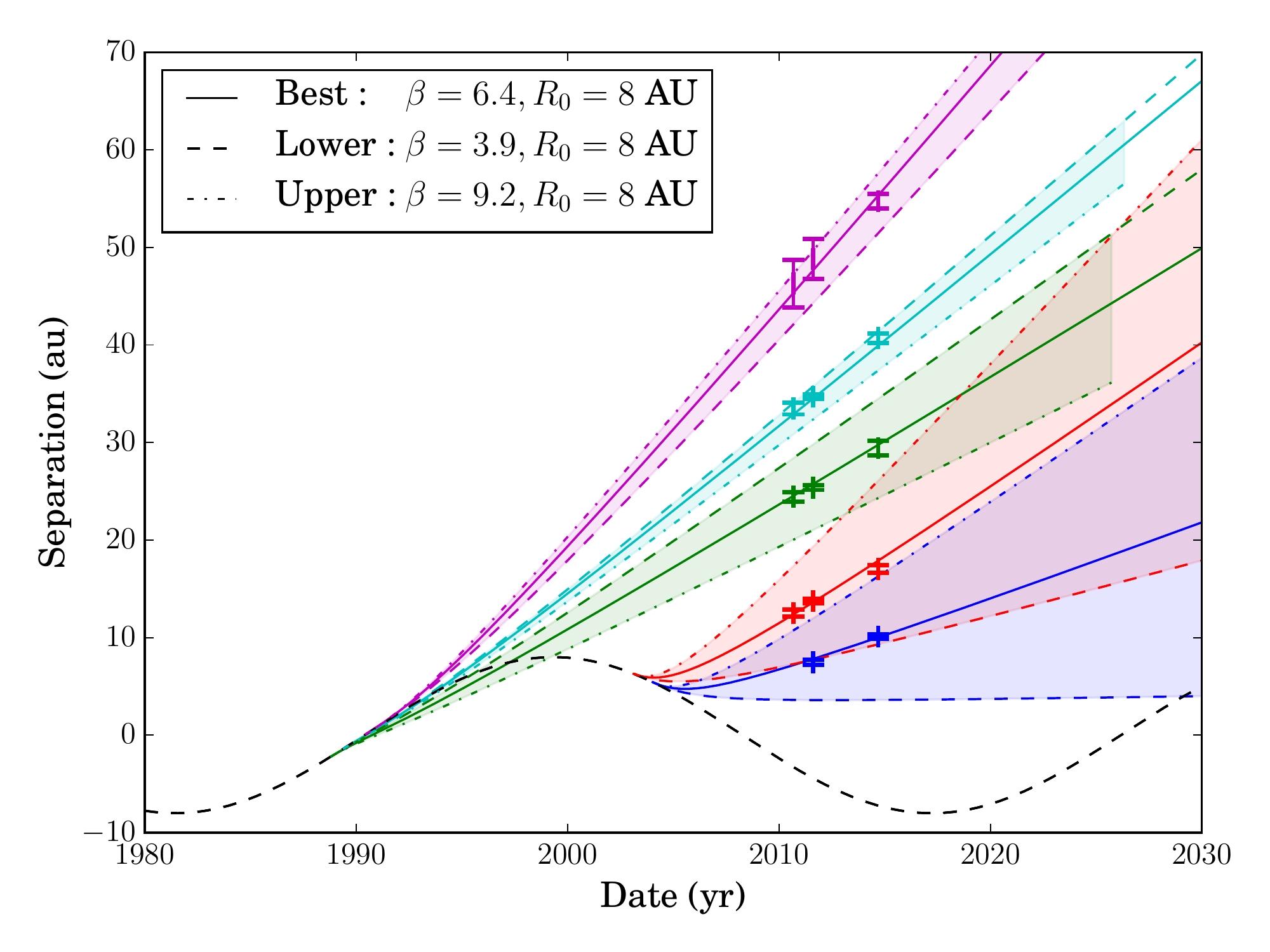}
}
\parbox{0.5\textwidth}{
\includegraphics[width=0.51\textwidth, height=!,trim=0 0.5cm 0  0.5cm,clip]{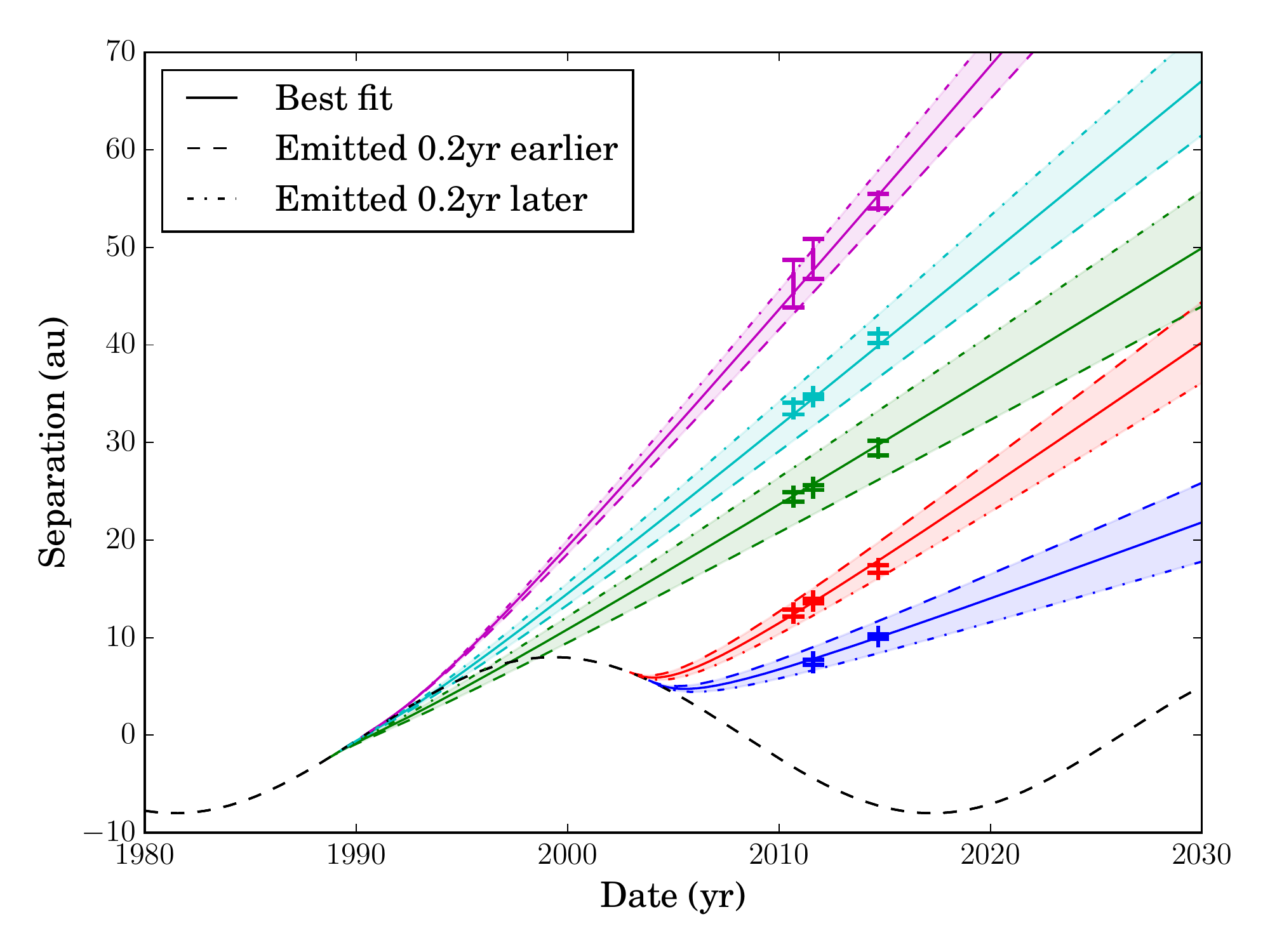}
}
}
\caption{\it Orbiting parent body. \rm \textbf{Left}:  Spatial extent
  of the arches if they were composed of particles with different
  \be\ values emitted at the same date. \textbf{Right}: Spatial extent of the structures if they were
  composed of particles released continuously during 0.4 year. The
  dashed black line corresponds to the apparent position of the parent
  body.}
\label{figDeltaBetaDate}
\end{figure*}

\subsection{Critical assessement of the model}

To assess the robustness of the model results, we evaluate in the
following the impact of some assumptions on our findings. 

\subsubsection{Consistency with the 2004 observations}
\label{sec:2004}

We have so far ignored the 2004 measurements since the arch-like
features are not detected as such in these data sets. The presumed
2004 locations of the D and E features documented in
Table~\ref{tabPos} correspond to reported positions of brightness
maxima in the literature rather than maximum elevations. On the other
hand, these data greatly increase the time base and this provides an
opportunity to check if the brightness maxima identified in 2004 would
be consistent with the dynamics of the D and E features that we
inferred.
We derived best fits to the 2004, 2010, 2011 and 2014 data altogether
by considering the case of an orbiting parent body, with no
restriction on the period of emission, a situation similar to the
nominal case in Sec.~\ref{subsecNomOrb}. The likeliest values of
\be\ and \R\ are reported in Figure~\ref{figBetaRaySol} for comparison
with those obtained previously. We find that adding the 2004
observations reduces the error bars but has a marginal impact on the
best values of the parameters (model labelled "2004 2014" in the
figures). 
 The best fit is indeed obtained for $\beta =
7.0^{+2.0}_{-1.8}$ and $R_0 = 8.2^{+0.9}_{-1.2}$~au. 
This compares
well with the values derived from the best fit to the 2010--2014 data
set, and introducing new data to the fit only yields a small increase
of the reduced \XX\ (2.4 \it vs \rm 1.7).
Therefore, we conclude that the 2004 brightness assymmetries in the
2004 images can be associated with structures D and E, as proposed in
\cite{Boccaletti2015}.
A more appropriate evaluation of these features will be presented in Boccaletti et al. (in prep.).

\subsubsection{Stellar parameters}
\label{sec:mass}

A parameter that can affect the modeling results is the stellar
mass. The uncertainty on the estimation of AU Mic's mass leads us to
examine the impact of an heavier star. We consider again the case of
an orbiting parent body with no constraint on the emission dates as in
Sec.~\ref{subsecNomOrb}, and we change the stellar mass from
0.4~M$_{\odot}$ to 0.7~M$_{\odot}$ (labelled "0.7 M$_{\odot}$" in
Fig.~\ref{figBetaRaySol}). 
 The best value of $\beta$ is essentially
not affected ($5.4^{+4.3}_{-2.5}$), but $R_0$ is increased to $9.7^{+1.6}_{-2.0}$~au such
that the orbital period is kept nearly constant with respect to the
case of a lower stellar mass.  In the 0.7~M$_{\odot}$ case, the parent
body has an orbital period of 36.1~years, against 33.8~years in the
solution of Section~\ref{subsecNomOrb}. 
It means that the time interval
between each dust release event is most significant than the radius of
emission. Overall, it shows that the uncertainty on the mass of the
star does not significantly impact our main conclusions.

 Another stellar parameter that can affect the simulations is the stellar wind speed,
here assumed to be equal to the escape velocity at the surface of the star, 
following the approach by \cite{Strubbe2006} and \cite{Schuppler2015}.
Although observations \citep{Luftinger2015} and models \citep{Wood2015} 
exist for the stellar wind of main-sequence solar-like stars of various ages, 
the constraints are very scarce for an active, young M-type star like AU~Mic.
In the literature, the values are either computed based on the escape velocity, 
or by considering the temperature at the base of the open coronal field lines together with the Parker's hydrodynamical model (\citeyear{Parker1958}).
This leads to wind speed values that can vary by a factor of up to 3 from a model to another.
We have numerically checked that multiplying by a factor of 10 our adopted value of 450~km/s for the wind speed
only changes marginally the dust dynamics.
This is true as long as the dust speed is neglectible with respect to the wind speed (see Sec. \ref{subsecPressureForce}).
As a consequence, the inferred best \be\ and \R\ parameters are not affected by the exact $V_{\rm sw}$ value assumed in the model.
However, the connection between \be\ and the grain size depends on the wind speed, as discussed in Sec.~\ref{sec:dustproperties}.

\subsubsection{$\beta$ distribution and event duration}
\label{subsecDistrib}

Our model intrinsically assumes that the observed features labelled A
to E are made of grains of a single size (unique \be\ value) and that
the dust release events are sufficiently short in time to be
considered as  instantaneous.  
In the case of an orbiting parent body, the
best fit value for $\beta$ shows a $1\sigma$ uncertainty of about
30\%.  That suggests a limited dispersion in \be\ values.  This is
illustrated in the left panel of Figure~\ref{figDeltaBetaDate} where
it is shown that $\Delta \beta / \beta$ of about 1/3 can be tolerated
as long as the D and E features are concerned, but the model becomes
increasingly inconsistent with the observations when considering the
features located closer and closer to the star. 
For the A, B, and C structures, we observe an overlap which would connect the features, contradicting the observations.
 This very much
suggests that either the arch-likes features are formed of grains with
a narrow size distribution, and/or that their cross sectional area is
dominated by grains in a narrow size range (see also
Sec.~\ref{sec:dustproperties}).

Likewise, assuming for example that the dust release events last a few
months significantly widens the range of apparent trajectories as
illustrated in Figure~\ref{figDeltaBetaDate} (right panel, $\Delta t =
\pm 0.2$~year).  However, this behaviour is still compatible with the
observations, since the structures are not mixed together, and have a
radial extent compatible with the one obtained here.  This suggests
that the emission process can occur during a few months, as long as it
stay shorter than the time difference between two consecutive
structures (0.6~year in this case).

Therefore, the on-going follow up on this system will be critical to further constrain the \be\ distribution and the duration of release events.

\subsection{Grain size and mass loss rate}
\label{sec:dustproperties}

\begin{figure}
\includegraphics[width=0.49\textwidth, height=!,trim=0 .5cm 0 .5cm,clip]{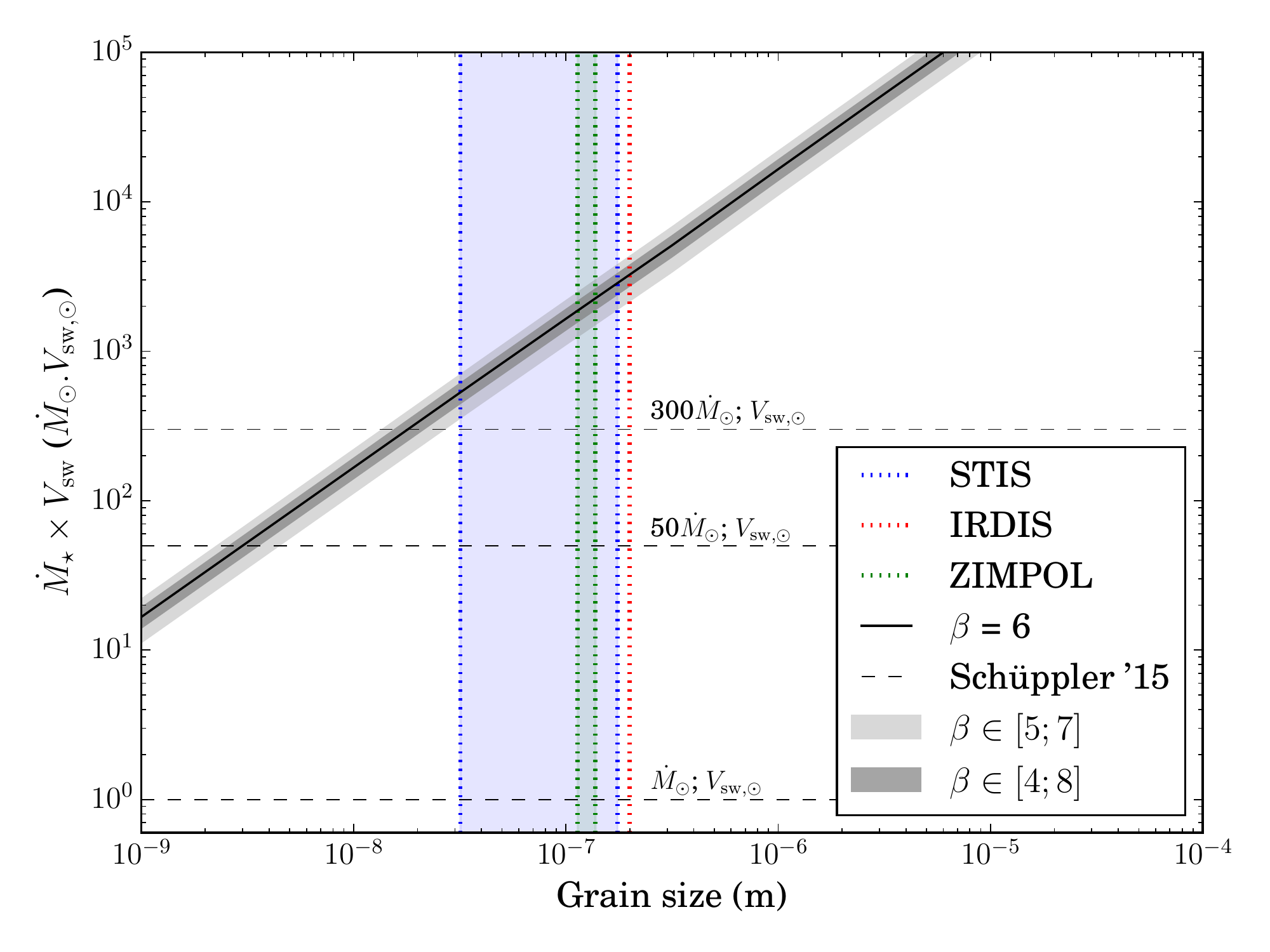}
\caption{Product of the mass loss rate with the wind speed \it vs \rm grain size for
  the likeliest values of \be\ obtained in our simulations.  
  The gray areas correspond to the
  dispersion of \be\ in the orbiting cases.
  For simplicity, our assumed value of wind speed 
  of 450~km/s is labelled $V_{\rm sw, \odot}$.
   The colored vertical lines indicate, for each observing wavelength, 
  the range of grain sizes corresponding to the smallest grains 
  that efficiently scatter light ($s \sim \lambda / 2 \pi$ ) 
  and which should dominate the flux for any "non-exotic" grain size distribution in the clumps 
  (see Sec.~\ref{sec:dustproperties} for details).  }
\label{figSizMas}
\end{figure}

 Our best fit value for \be\ (about 6.3 in the case of a free orbiting
parent body)  
is large enough to consider that the contribution of the
radiation pressure to the dynamics of the grains forming the arch-like
features can be neglected. Indeed, the low luminosity of the star
makes $\beta_{\rm PR}$ never exceeding 0.3 as can be seen in Fig.~10
of \cite{Augereau2006}.  Therefore, we can assume $\beta \simeq
\beta\dma{SW}$, 
 and as a consequence, the link between \be\ and the
grain size $s$ is degenerated with the mass loss rate
$\dot{M}_{\star}$ and the stellar wind speed $V_{\rm sw}$, such that $\beta \propto \dot{M}_{\star} V_{\rm sw} / s$
(Eq.~\ref{eqBsw}).
This is illustrated in Figure~\ref{figSizMas},
using dust composition M1 of \cite{Schuppler2015} ($\rho$ = 1.78~g.cm$^{-3}$, see their Tab.~2). 

  In this context it is interesting to question which grain sizes are probed by the visible/NIR scattered light observations.
For the purpose of the discussion, we can approximate the
dimensionless scattering efficiency $Q\dma{sca}$ by a constant for
grains much larger than the observing wavelength (geometric optics, $x\gg 1$ where $x = 2\pi s/\lambda$ is the size parameter), 
and $Q\dma{sca} \propto x^4$ for small grains in the Rayleigh regime ($x\ll 1$).  
The differential scattering cross section, that writes $Q\dma{sca} \pi s^2 \ma{d}n(s)$, 
is proportional to $s^{6+\kappa}\ma{d}s$ for $x\ll 1$
and $s^{2+\kappa}\ma{d}s$ for $x\gg 1$
when considering a power law differential grain size distribution $\ma{d}n \propto s^{\kappa}\ma{d}s$ with a lower cut-off size $s\dma{min}$. 
For any value of $\kappa$ such as $-7<\kappa<-3$, 
as for instance the classical collisional "equilibrium" size distribution in $\kappa=-3.5$
\footnote{Note, however, that such an equilibrium distribution might not apply across the \be=0.5 limit, where a sharp transition is expected \citep{Krivov2010}.}, 
the scattering cross section will be dominated by grains such as  $s\sim \lambda/2\pi$ (i.e. $x\sim 1$).
The ranges of grain sizes that these
correspond to are displayed in Figure~\ref{figSizMas} for the HST/STIS
(broad-band, 0.2--1.1\,$\mu$m), SPHERE/IRDIS (J-band) and
SPHERE/ZIMPOL (I'-band) observations, 
 and are typically of the order of $\sim 0.1\,\mu$m.
In order for grains of this size to reach our likeliest \be\ value of $\sim 6$,
we need the  $\dot{M}_{\star} \times V_{\rm sw}$ to reach values 
as high as a few $10^3$ the solar analog.
Such values are at least 20 times greater than the 50\,$\dot{M}_{\odot} \times V_{\rm sw, \odot}$
derived by \citet{Schuppler2015} from collisional modelling of the overall disk.
However, such very large values cannot be fully ruled out because there is a large spread
of $\dot{M}_{\star} V_{\rm sw}$ estimates reported in the literature
for M-type stars, including values 
3 to 4 orders of magnitude larger than the solar case
\citep[see e.g.][and references therein]{Vidotto2011}. 
The global trends of $\dot{M}_{\star}$ decreasing with age and increasing with stellar activity 
\citep{Wood2005} favor a high mass-loss rate in the case of AU~Mic.

It remains to be checked, however, whether the apparent positions and
velocities are sensitive to the observing wavelength, which is
difficult to conclude with current data because the spectral range of
the observations is limited.  With a collisional grain size
distribution, one would indeed expect that the structures at visible
wavelengths might be formed of smaller grains with larger \be\ values
than the structures observed in the near-infrared (smaller
\be\ values).  
 An alternative would be that the size distribution is very narrow, 
which can be schematically described by a steep size distribution 
with a minimum size cutoff.  
For $\kappa < -7$,
the scattering cross section is always dominated by the smallest grains of
the size distribution  ($s\dma{min}$), 
regardless of the observing wavelength.  
In this case, the features'  measured positions and velocities would be the same at all wavelengths, 
and all images could be dominated by the same grain sizes, 
which could be much smaller than 0.1\,$\mu$m 
 and thus requiring relatively moderate
$\dot{M}_{\star} V_{\rm sw}$, typically 10$^2$ the solar value. 
This would be consitent with the blue color of the overall disk,
indicating a cross sectional area dominated by submicrometer-sized grains,
while micrometer-sized grains would produce gray scattering
\citep{Augereau2006,Fitzgerald2007,Lomax2017}.

 We conclude that either the features are formed of grains with a size distribution that is consistent with being collisional,
thus requiring a high stellar mass-loss rate;
or that they are formed of grains in a very narrow range around very small sizes 
($\ll 0.1 \mu$m) allowing moderate mass-loss rate 
but requiring a physical explanation for the presence 
 of such a large amount of nano-grains 
and the relative absence of slightly bigger grains 
(because the size distribution is extremely peaked around~$s\dma{min}$).

\subsection{Detected and undected features}

\subsubsection{Future positions of observed features and parent body}

\begin{figure}
\includegraphics[width=0.49\textwidth, height=!,trim=0 0.5cm 0 0.5cm,clip]{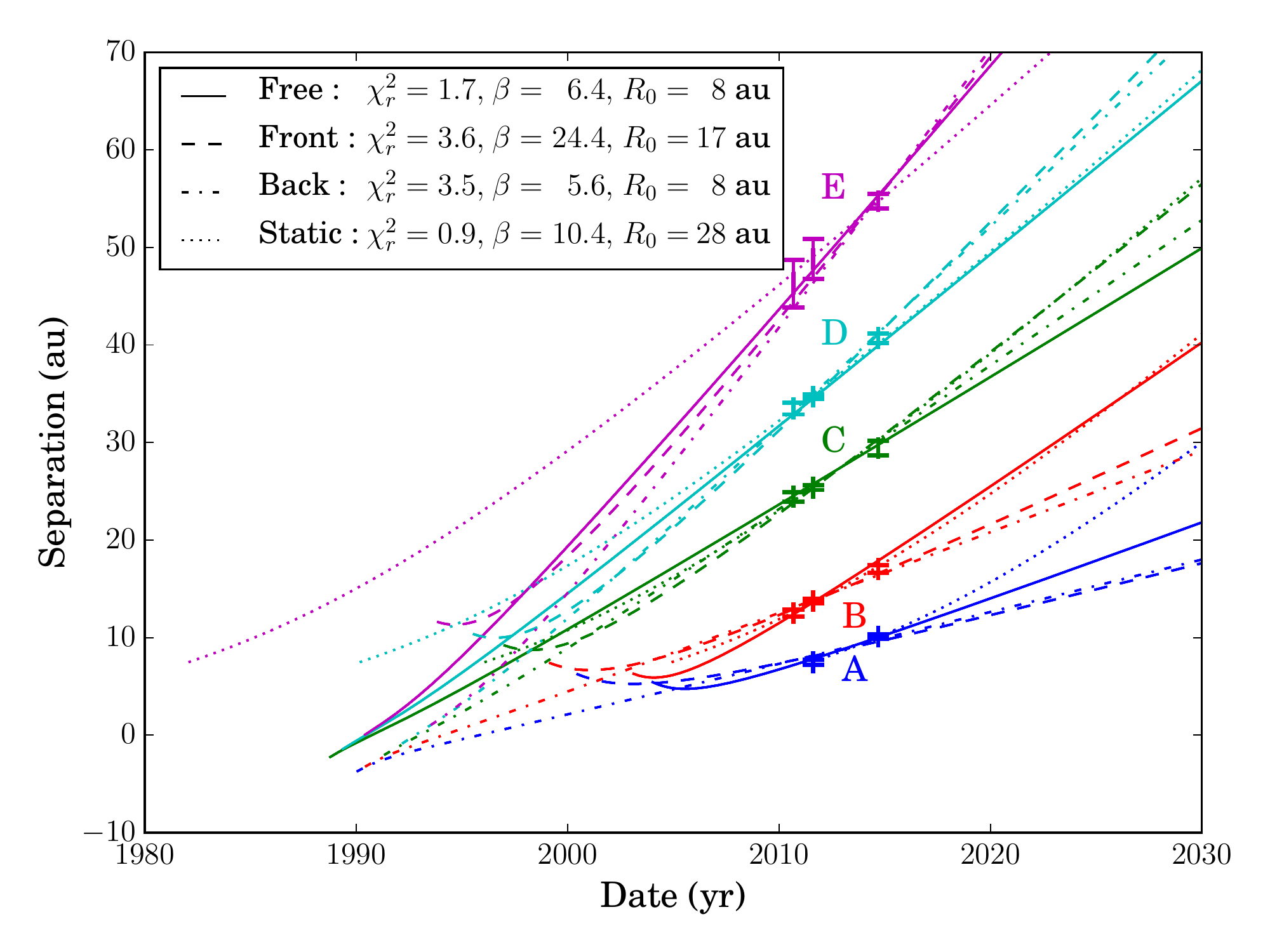}
\caption{Apparent positions of the features as a function of the
  observing date for the 4 optimal cases discussed in
  Sec.~\ref{secResults}.}
\label{figPosComp}
\end{figure}

Our model yields constraints onto the spatial and temporal origin of
the grains forming the fast-moving features. This can be used to
predict the future positions of the structures and offers an
opportunity to better isolate, with upcoming observations, a best
scenario out of the four discussed in this study and summarized in
Fig.~\ref{figPosComp}. 
Nevertheless, this figure clearly shows that the differences in apparent positions of the features 
according to the various scenarii start to become significant 
at least a few years after the most recent data used in this paper.

The predicted positions of the features for each model are
documented in Table~\ref{tabPredic}. In 2020 for instance, the
predicted positions differ by typically a few au (a few 0.1$\arcsec$)
which is in principle large enough to reject some of the proposed
scenarii. However, we warn that these plausible positions of the
features are idealized and do not take into account the uncertainties
on the model parameters. In summary, the apparent trajectories of the
known structures need to be followed in time and can be compared to
our model predictions, but this might not be enough to identify within
the next few years the most realistic scenario among the four
presented in this paper.

Interestingly, we note that, 
 if its orbit is exactly seen edge-on, 
the unseen parent body should have transited, 
or will at some point transit in front of the star. 
In all the parent body orbiting models, 
we expect it to have transited during the 2000--2014 time period, 
if its orbit is anti-clockwise. 
 The free orbiting case predicts that the transit occured in 2008.5, 
the frontward orbiting case predicts it in 2007.1 
and the backward orbiting case in 2010.7. 
Light curves of the star taken during this period could evidence this hypothetic transit 
(although AU Mic is active). 
If the orbit is
clockwise, the next transit is planned in 2026.4 in the free case, in
2062.5 in the frontward case, and in 2028.6 in the backward case.

\begin{table*}
\caption{Prediction of position (in arcsecond) for the five structures.
The uncertainties correspond to the dispersion due to the reference structure chosen.}
\label{tabPredic}
\begin{center}
\begin{tabular}{| c| l| c|  c|  c|  c|  c |}
\hline
Date	& Model & A		& B		& C		& D		& E \\
\hline
\multirow{4}{*}{2015.0} & Static & 1.04 $\pm$ 0.01 & 1.78 $\pm$ 0.01 & 3.09 $\pm$ 0.03 & 4.09 $\pm$ 0.04 & 5.55 $\pm$ 0.02 \\
 & Orbiting free & 1.03 $\pm$ 0.01 & 1.83 $\pm$ 0.02 & 2.97 $\pm$ 0.04 & 4.08 $\pm$ 0.03 & 5.63 $\pm$ 0.03 \\
 & Orbiting frontward & 0.98 $\pm$ 0.01 & 1.69 $\pm$ 0.01 & 3.10 $\pm$ 0.02 & 4.20 $\pm$ 0.02 & 5.62 $\pm$ 0.01 \\
 & Orbiting backward & 0.99 $\pm$ 0.03 & 1.66 $\pm$ 0.04 & 3.05 $\pm$ 0.01 & 4.19 $\pm$ 0.02 & 5.65 $\pm$ 0.01 \\
\hline
\multirow{4}{*}{2017.0} & Static & 1.24 $\pm$ 0.02 & 2.05 $\pm$ 0.01 & 3.42 $\pm$ 0.05 & 4.44 $\pm$ 0.07 & 5.93 $\pm$ 0.06 \\
 & Orbiting free & 1.18 $\pm$ 0.02 & 2.11 $\pm$ 0.03 & 3.23 $\pm$ 0.05 & 4.44 $\pm$ 0.05 & 6.13 $\pm$ 0.04 \\
 & Orbiting frontward & 1.08 $\pm$ 0.01 & 1.88 $\pm$ 0.01 & 3.44 $\pm$ 0.02 & 4.63 $\pm$ 0.02 & 6.17 $\pm$ 0.02 \\
 & Orbiting backward & 1.10 $\pm$ 0.03 & 1.83 $\pm$ 0.04 & 3.35 $\pm$ 0.01 & 4.60 $\pm$ 0.04 & 6.23 $\pm$ 0.02 \\
\hline
\multirow{4}{*}{2020.0} & Static	 & 1.58 $\pm$ 0.03 & 2.49 $\pm$ 0.01 & 3.93 $\pm$ 0.08 & 4.98 $\pm$ 0.12 & 6.49 $\pm$ 0.12 \\
 & Orbiting free	 & 1.41 $\pm$ 0.04 & 2.55 $\pm$ 0.06 & 3.62 $\pm$ 0.06 & 4.98 $\pm$ 0.08 & 6.90 $\pm$ 0.06 \\
 & Orbiting frontward & 1.24 $\pm$ 0.01 & 2.17 $\pm$ 0.02 & 3.95 $\pm$ 0.03 & 5.29 $\pm$ 0.03 & 6.99 $\pm$ 0.03 \\
 & Orbiting backward & 1.26 $\pm$ 0.04 & 2.08 $\pm$ 0.04 & 3.80 $\pm$ 0.03 & 5.23 $\pm$ 0.06 & 7.11 $\pm$ 0.05 \\
\hline
\multirow{4}{*}{2025.0} & Static	 & 2.25 $\pm$ 0.04 & 3.28 $\pm$ 0.04 & 4.82 $\pm$ 0.16 & 5.91 $\pm$ 0.21 & 7.45 $\pm$ 0.53 \\
 & Orbiting free	 & 1.81 $\pm$ 0.07 & 3.30 $\pm$ 0.11 & 4.28 $\pm$ 0.07 & 5.89 $\pm$ 0.12 & 8.18 $\pm$ 0.11 \\
 & Orbiting frontward & 1.50 $\pm$ 0.01 & 2.66 $\pm$ 0.02 & 4.81 $\pm$ 0.04 & 6.39 $\pm$ 0.05 & 8.38 $\pm$ 0.04 \\
 & Orbiting backward & 1.53 $\pm$ 0.04 & 2.50 $\pm$ 0.05 & 4.54 $\pm$ 0.05 & 6.26 $\pm$ 0.09 & 8.58 $\pm$ 0.10 \\
\hline
\end{tabular}
\end{center}
\end{table*}

\begin{figure*}[tp!]
\centering
\hbox to \textwidth
{
\parbox[b]{0.33\textwidth}{
\includegraphics[width=0.335\textwidth, trim=4.5cm 0.6cm 6.5cm 0.3cm, clip]{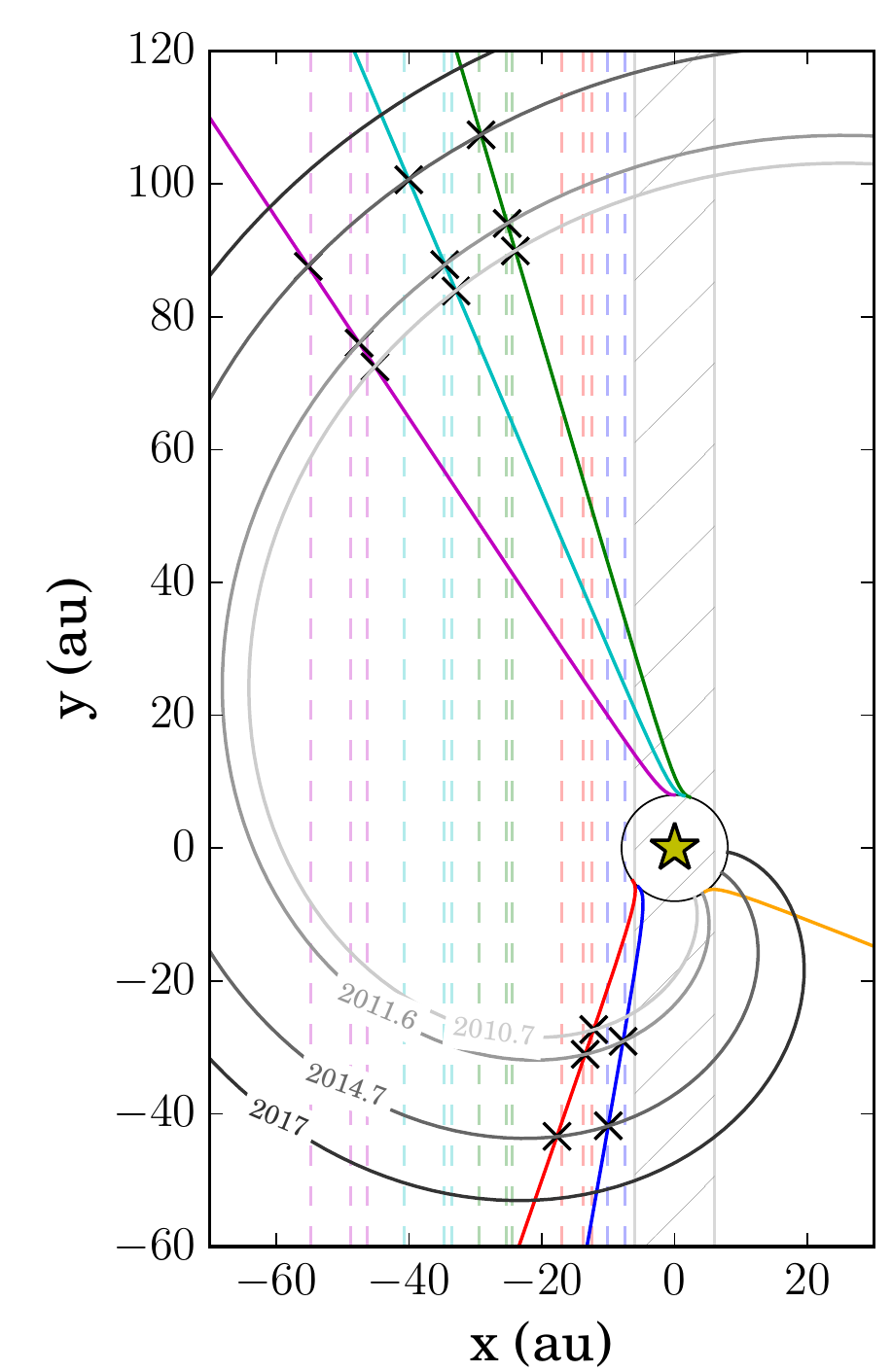}
\subcaption{Model without constraint}
}
\parbox[b]{0.33\textwidth}{
\includegraphics[width=0.33\textwidth, trim=4.5cm 0.6cm 6.5cm 0.3cm, clip]{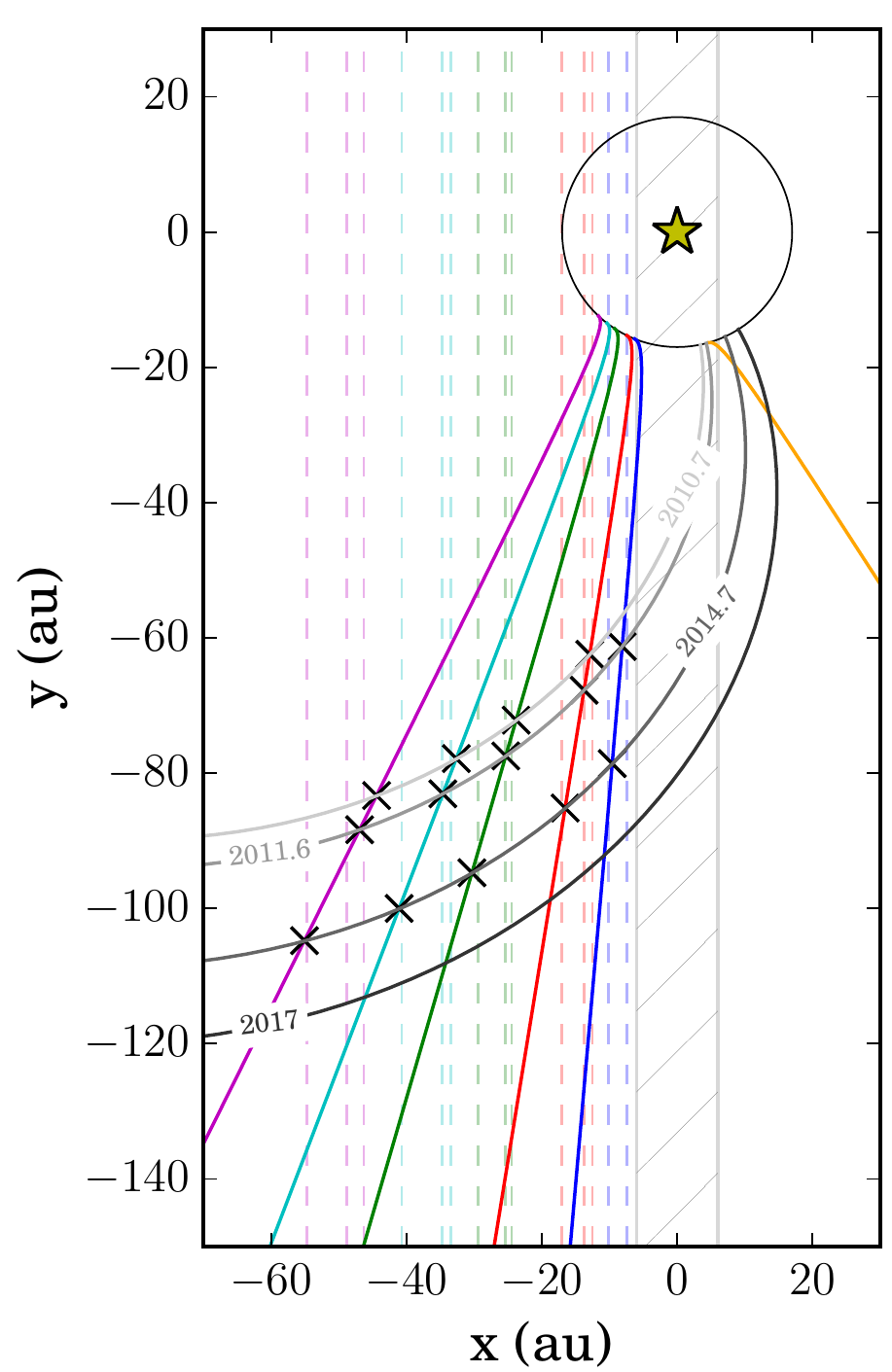}
\subcaption{Particles going toward the observer}
}
\parbox[b]{0.33\textwidth}{
\includegraphics[width=0.33\textwidth, trim=4.5cm 0.6cm 6.5cm 0.3cm, clip]{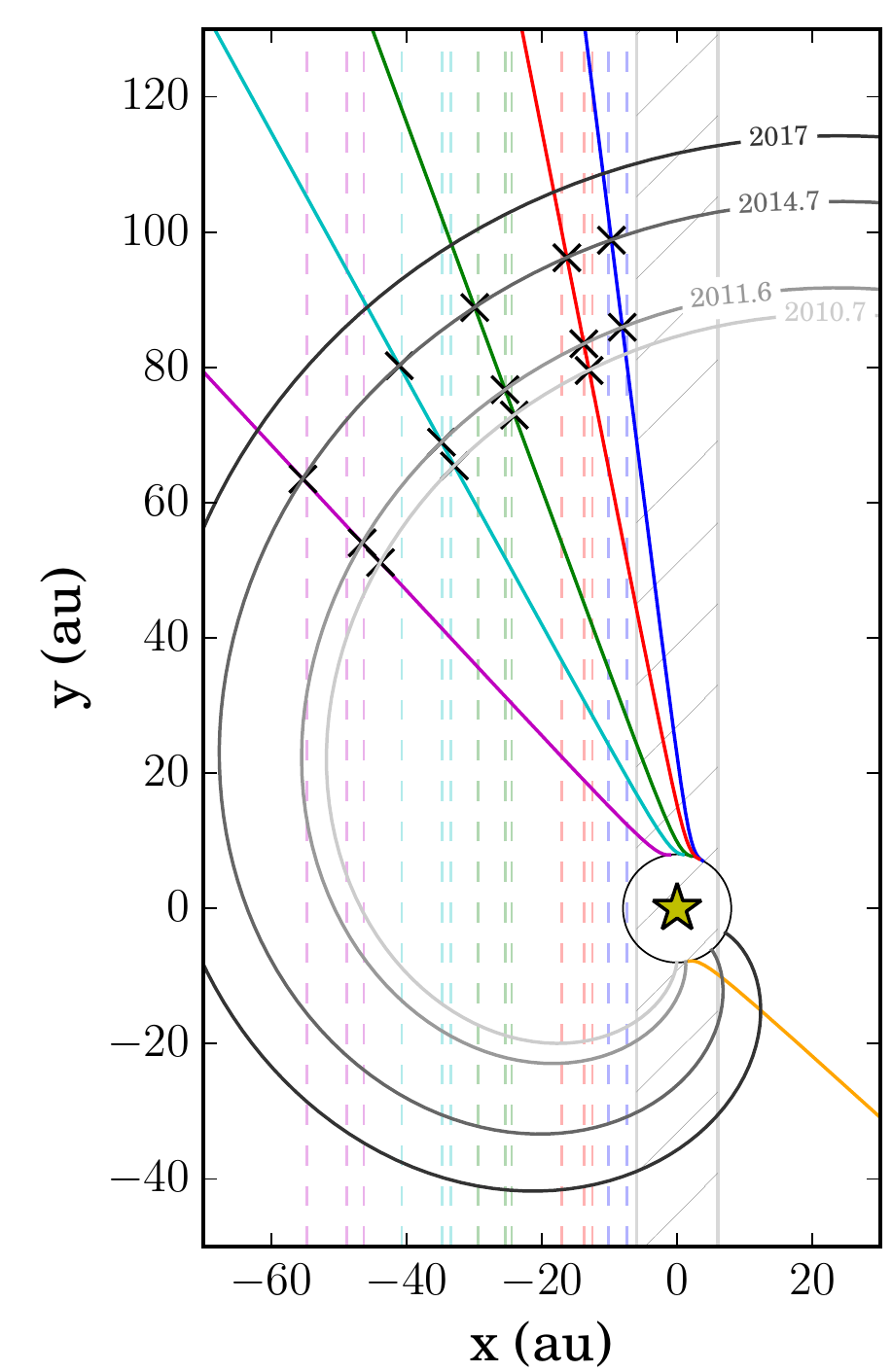}
\subcaption{Particles going away from the observer}
}
}
\caption{\it Orbiting parent body. \rm
Trajectories of the particles seen from above averaged over the five
  reference structures (see Secs.~\ref{subsecNomOrb} and \ref{sec:grouped} for details).
  The color-coding is the same as in
  previous figures. 
The orange trajectory is the one of fictive particles released in 2012.  
  The solid black circle is the trajectory of the parent body. The crosses correspond to the observing dates. 
  The gray lines are the position of the dust in 2010.7, 2011.6, 2014.7 and 2017 (from the fainter to the darker), if it was continuously emitted by the
  parent body on its orbit.
  The hatched area roughly corresponds to the masked inner region in case of STIS imager.}
\label{figProspec}
\end{figure*}

\subsubsection{Missing and future features}

The observed structures are recent, and the dust release events occurred several times over the last 25--30 years. Some scenarii are characterized with a pseudo periodic behaviour, which allows us to predict new structures to appear.
Therefore,  it is crucial to determine whether some additional features will be, or should have been, detected.
In the static case for
instance, we expect any new feature to be localized on the same side
of the disk (southeast), and to be emitted with an about 7-year
periodicity (Sec.~\ref{subsecNomSta}). This suggests that the next
feature in the static case would be emitted in between 2018 and 2020,
and the evolution of its apparent position and projected speed should
be similar to feature A, but shifted in time by about 7 years. Perhaps
more importantly, should any new structure be detected on the
northwest side of the disk, the static parent body model would
immediately be discarded.

To follow hypothetic new structures in the case of an orbiting parent
body, we overplot in Figure~\ref{figProspec} snapshots of the dust
grain positions at specific observing dates as if they were emitted
continuously from the parent body.
In the parent body rotating frame, these positions
would correspond to streaklines.
At a given observing date, any previously released structure should be
located on this line. 
In the parent body orbiting model with ungrouped released events, the
model suggests a low or non-activity period between about 1990 and
2000. Structures that would have been emitted during this time period
would be located between 20 and 90~au in apparent separation from the
star, on the southeast side of the disk.  Structures emitted after the
most recent feature (feature A in about 2004) would have been too
close from the star until 2012 to be detected with available
instrumentation (panel~(a) of Figure~\ref{figProspec}). After this
date, new structures would be observable on the northwest side and
would have been seen with VLT/SPHERE in 2014. Their non detection
could for instance suggest that the system entered a similar inactive
period as in 1990-2000.

In the scenario where all structures are moving
away from the observer, the most recent structure (feature E) has been
emitted late 1993, and new structures possibly emitted during the
1994--2010 time period would be observable on the southeast side of
the disk, as shown on panel (c) of Fig.~\ref{figProspec}. Their
non-detection suggests either the process of dust release has stopped
for at least 15 years, or the dust release process is much less
efficient during that time interval, making the structures not
observable (too faint, for example).  Following the 2014 streakline,
we also notice that no structure could be located further than 70~au
from the star in apparent separation in the images used in this
study. The features possibly emitted after about 2010 would have been
too close to the star to be detected until now, and this model
predicts that new features could become observable on the northwest
side of the disk in upcoming observations.

In the case of an orbiting parent body with grouped emissions toward
the observer, the oldest structure was emitted in late 1993. Older features
would be located beyond the E structure in projection, and could have
been too faint to be detected. Therefore, this model is consistent
with the lack of more distant features in the HST/STIS images (the VLT/SPHERE field of view is limited to 6''), despite a
possible 1.5 to 2 year pseudo-periodicity (Sec.~\ref{sec:grouped}).
The model also suggests that the most recent structure (feature A) is
emitted in 2000, and panel~(b) of Figure~\ref{figProspec}
shows that any feature formed during the 2000--2010 time period would
be essentially lying along the line of sight to the star, yielding
very small projected separations, preventing their detection with the
VLT/SPHERE and HST/STIS images used in this paper. Therefore, the
parent body could have continued emitting periodically since 2000
while remaining consistent with the non-detection of additional
features.  We note however that GPI observations by \citet{Wang2015}
identified a source possibly corresponding to a compact clump of dust,
within the apparent position of the A feature, and that would be
consistent with a new structure emitted in 2001.1. Structures possibly
emitted after 2011 should have been observed on the northwest side of
the disk in 2014 (see for example the orange trajectory for the
position of hypothetic structures arbitrarily emitted in 2012). This
suggests again that either the pseudo-periodicity is too loose to
predict precisely the arrival of future structures, or that the
emission process has stopped. We note, however, that this orbiting
parent body model remains the most consistent with a periodic
behaviour and the lack of detected features on the northwest side of
the disk so far.
Here again, a systematic monitoring is the key to address the actual evolution of the system.


\section{Conclusion}

We construct a model to reproduce the apparent positions of the
structures observed in the debris disk of AU Mic, 
taking into account the
stellar wind and radiative pressure onto the dust grains,
 assuming that we observe the proper motion of the dust.
We do not investigate the possible physical process at the origin of the dust production, 
but consider two different dynamical configurations for the release of the observed dust: 
a common origin from a fixed location static with respect to the observer, 
or release from an hypothetical parent body on a Keplerian orbit.  
In all cases, we find that the dust seems to originate from inside the planetesimal belt,
at typically 8~au from the star in the best orbiting-parent-body model, 
or 28~au in the static case.  
The high projected velocities mesured for each structure require that the observed grains have a high
value of \be\ ($\sim 6$), the ratio of pressure and radiative forces on the
gravitational force. 
Our study could not disentangle between all the
scenarii considered based on the available observations.
However, we are able to predict, for each scenario, the future behaviour of the
structures and we discuss the hypothetic appearence of new
structures, especially on the northwest side of the disk. 
For all the scenarii, we find a semi-periodic behaviour of dust release.
We could also associate the brightness maxima observed in the 2004 images with
the fast-moving structures resolved in the more recent high-contrast images. 
 We suggest that the arch-like structures are 
either formed from $\sim 0.1~\mu$m-sized grains if the stellar wind is very strong, 
or from nanometer-sized grains ($\lesssim 20$~nm) with a very narrow size distribution, 
in the case of a more moderate stellar activity. 

Our model does not provide direct constraints on the source of dust
(parent body) nor on the circumstances that yield to a release event.
 We can however say that it must be somewhat periodic, 
and that every release event should last less than 6 months.
Furthermore, it must produce a great amount of submicron-sized grains, 
possibly with a narrow size distribution. 
Our static parent body model could correspond to planetesimals and
dust formed after a giant collision, while an orbiting parent body
could correspond to an unseen planet or a local concentration of dust
due to resonant trapping with a planet, for instance. 
A process of accretion onto the parent body, leading to ejection \citep[see e.g.][]{Joergens2013}
 can also be the origin of dust.
The stellar wind plays a key role in our model and it is likely that
the dust release events from the parent body are linked to the stellar
activity. The stellar flares themself are much too frequent to be the
triggering process responsible for the feature formation. We speculate
that this could be linked to the inversion of the magnetic field sign
of AU Mic, and could help forming arches 
 \citep{Sezestre2016, Chiang2017}. 
Overall, this model gives the base to a more complex model taking into
account the vertical elevation of the structures that we will address in a future paper.


\begin{acknowledgements} 
We thank Glenn Schneider and the HST/GO 12228 Team 
for the use of their reduced STIS image presented in Fig. \ref{imgBoc}.
We thank the referee,
Herv\'e Beust, Micka\"el Bonnefoy, Quentin Kral, the VLT/SPHERE consortium
team members, and the Exoplan\`etes team at IPAG for useful
comments that helped improving the paper. 
This work was supported by the "Programme National de Plan\'etologie" (PNP) of CNRS/INSU
co-funded by the CNES.
\end{acknowledgements}

\bibliographystyle{aa}
\bibliography{library}

\begin{appendix}

\section{Figures}

\begin{figure}[htp!]
\centering
\includegraphics[width=0.5\textwidth, height=!,trim=4.5cm 0.5cm 6cm 0.3cm,clip]{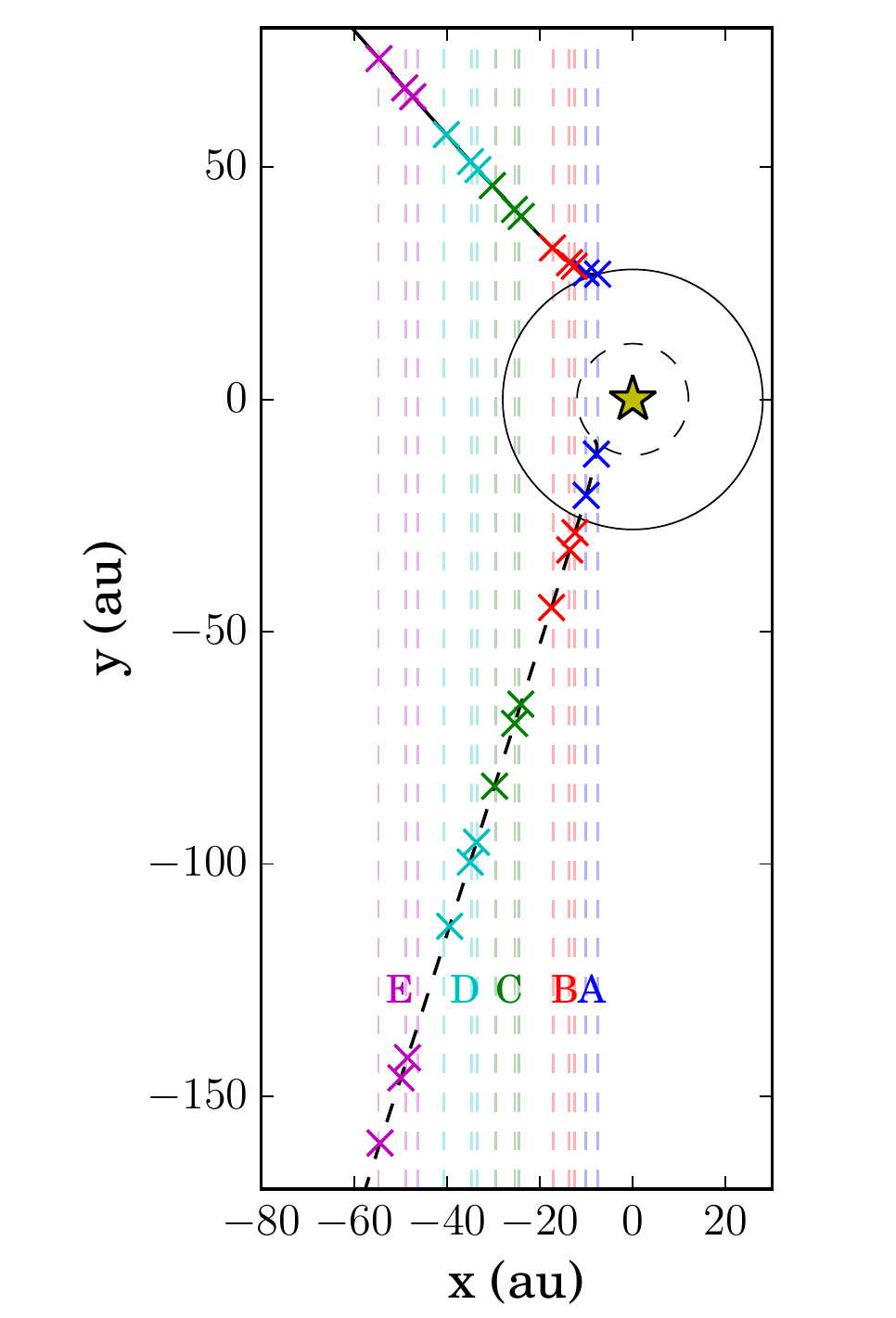}
\caption{\it Static parent body. \rm Trajectories of the particles
  seen from above that best fit the apparent positions (dashed lines,
  color-coding similar to Fig.~\ref{figSpeAdjABCDE}b,d) of five
  structures A, B, C, D and E in the case of a static parent body.  It
  corresponds to the trajectories plotted in black in
  Fig.~\ref{figSpeAdjABCDE}d, namely (\R , \be , $\theta$) = (28\,au,
  10.4, 165$\degr$) and (12\,au, 10.4, 43$\degr$). The line of
  sight of observer is assumed to lie along the $y$-axis, in the
  direction of increasing $y$ values, with the northeast side of the
  disk being on the left.}
\label{figAboStaABCDE}
\end{figure}

\begin{figure}[ht!]
\centering
\includegraphics[width=0.5\textwidth, height=!,trim=0 0.5cm 0 0.5cm,clip]{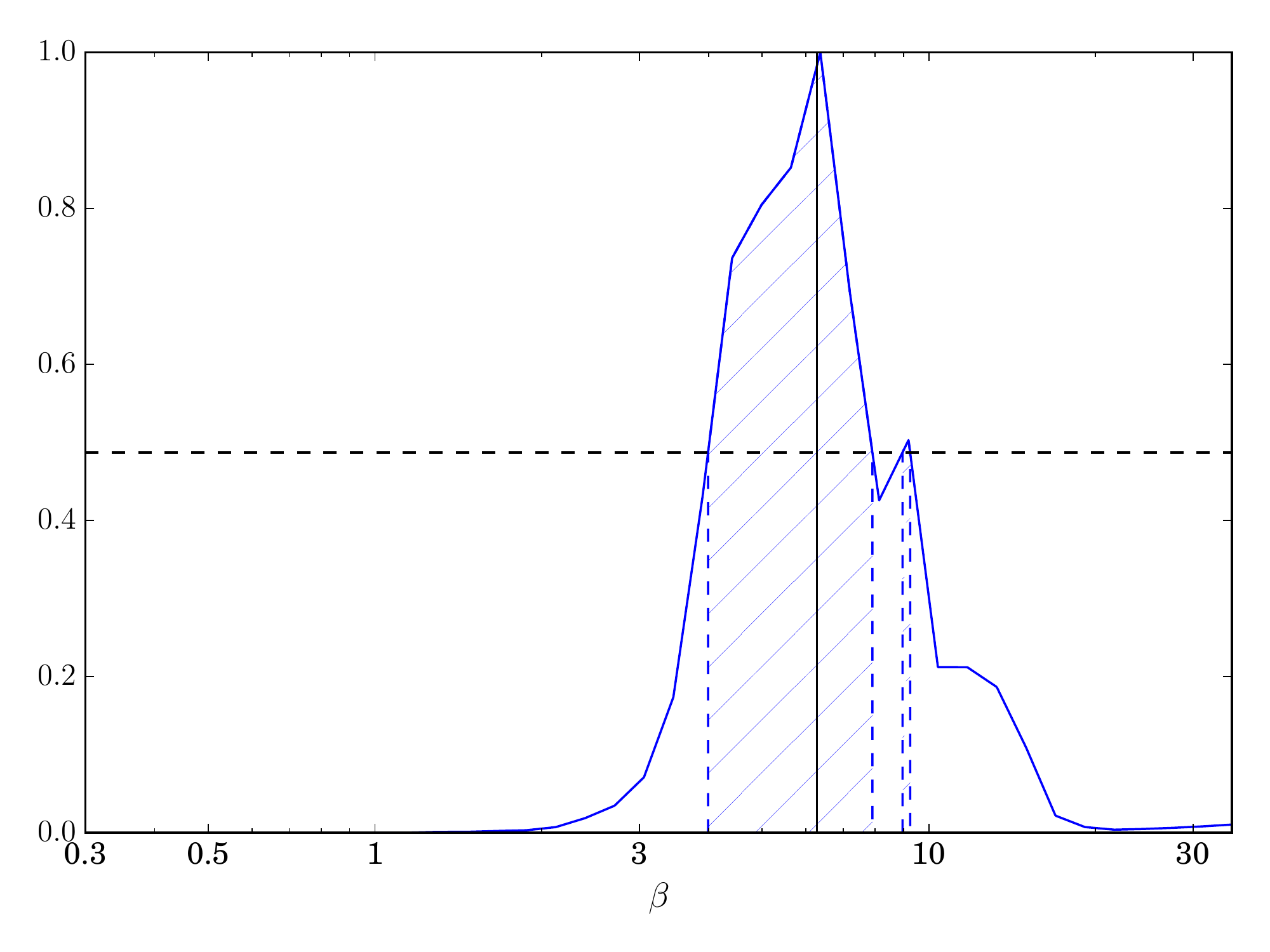}
\caption{\it Orbiting parent body. \rm 
Normalized probability distributions of \be.
The vertical black line is the mean value, and the dashed area corresponds to the 1$\sigma$ distribution.}
\label{figDistribBeta}
\end{figure}

\begin{figure}[ht!]
\centering
\includegraphics[width=0.5\textwidth, trim=0 0.5cm 0 0.5cm,clip]{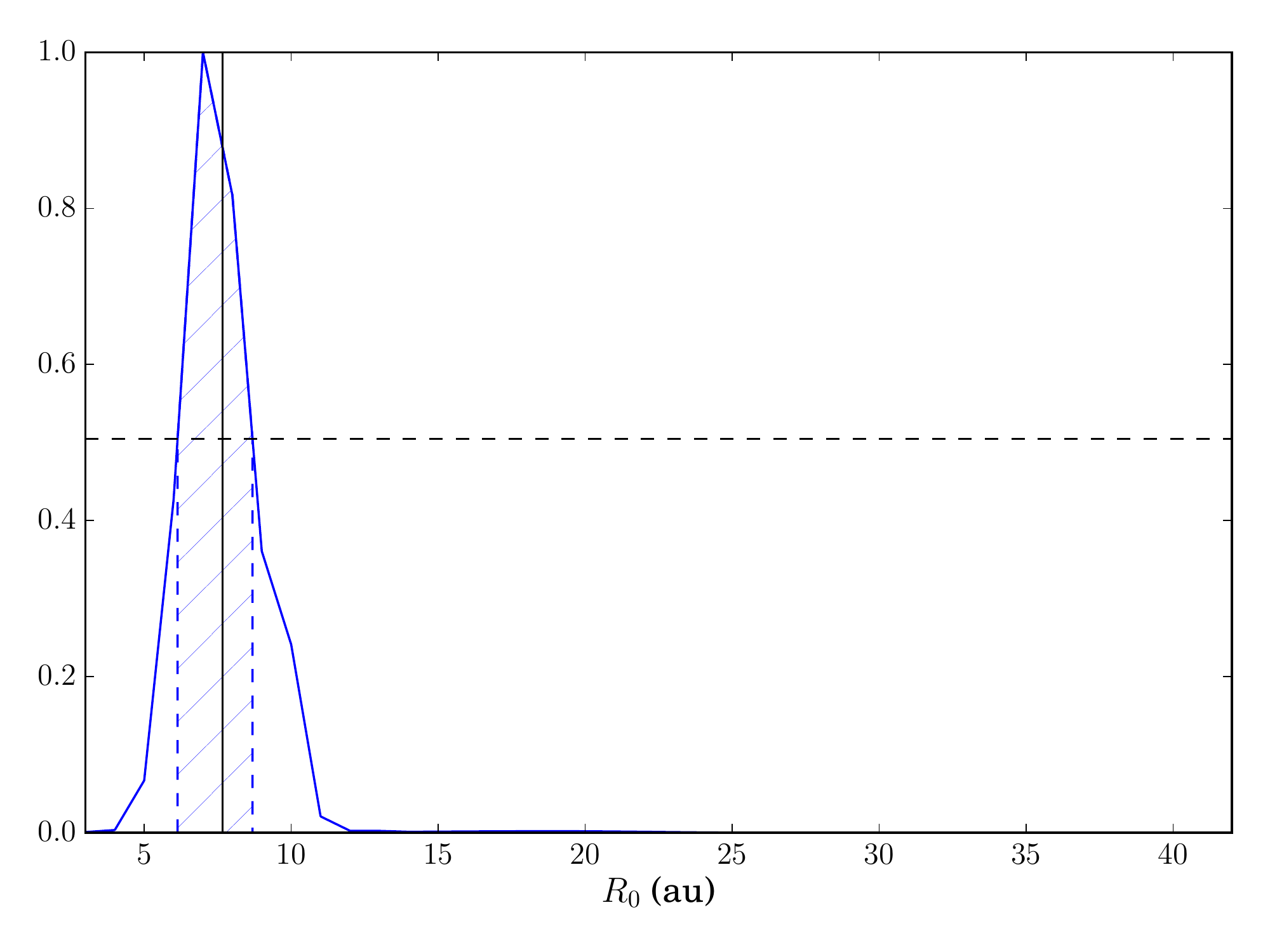} 
\caption{\it Orbiting parent body. \rm 
Normalized probability distributions of \R.}
\label{figDistribRay}
\end{figure}

\begin{table*}[h!]
\caption{Positions and speeds of the five structures (A to E) at
  different epochs derived from the data documented in
  Tab.~\ref{tabPos}. Any position corresponds to a mean value over
  time. The values in this table are plotted in
  Figure~\ref{figVitObs}.}
\label{tabValVit}
\begin{center}

\begin{tabular}{c c r r r r r}
\hline
Epoch & Variable & A & B & C & D & E \\
\hline
2004-2010 & x (au) & - & - & - & 29.01 & 39.15 \\
 & $\delta$x (au) & - & - & - & 0.58 & 0.90 \\
 & V (km/s) & - & - & - & 6.91 & 11.02 \\
 & $\delta$V (km/s) & - & - & - & 0.98 & 1.52 \\
\hline
2004-2011 & x (au) & - & - & - & 29.62 & 40.42 \\
 & $\delta$x (au) & - & - & - & 0.55 & 0.77 \\
 & V (km/s) & - & - & - & 6.81 & 11.26 \\
 & $\delta$V (km/s) & - & - & - & 0.81 & 1.17 \\
\hline
2004-2014 & x (au) & - & - & - & 32.62 & 43.38 \\
 & $\delta$x (au) & - & - & - & 0.57 & 0.36 \\
 & V (km/s) & - & - & - & 7.56 & 10.62 \\
 & $\delta$V (km/s) & - & - & - & 0.59 & 0.50 \\
\hline
2010-2011 & x (au) & - & 13.14 & 24.91 & 34.09 & 47.56 \\
 & $\delta$x (au) & - & 0.16 & 0.19 & 0.23 & 1.13 \\
 & V (km/s) & - & 6.29 & 4.78 & 6.14 & 12.79 \\
 & $\delta$V (km/s) & - & 1.59 & 1.96 & 2.35 & 11.44 \\
\hline
2010-2014 & x (au) & - & 14.78 & 26.94 & 37.10 & 50.53 \\
 & $\delta$x (au) & - & 0.18 & 0.31 & 0.27 & 0.90 \\
 & V (km/s) & - & 5.36 & 5.91 & 8.56 & 10.01 \\
 & $\delta$V (km/s) & - & 0.44 & 0.73 & 0.65 & 2.13 \\
\hline
2011-2014 & x (au) & 8.78 & 15.40 & 27.41 & 37.71 & 51.79 \\
 & $\delta$x (au) & 0.12 & 0.16 & 0.27 & 0.19 & 0.78 \\
 & V (km/s) & 4.10 & 5.07 & 6.25 & 9.30 & 9.16 \\
 & $\delta$V (km/s) & 0.38 & 0.49 & 0.84 & 0.60 & 2.40 \\
\hline
\end{tabular}
\end{center}
\end{table*}

\begin{figure*}[p!]
\centering
\hbox to \textwidth{
\parbox{0.5\textwidth}{
\includegraphics[width=0.5\textwidth, height=!,trim=0 0.5cm 0 0.5cm,clip]{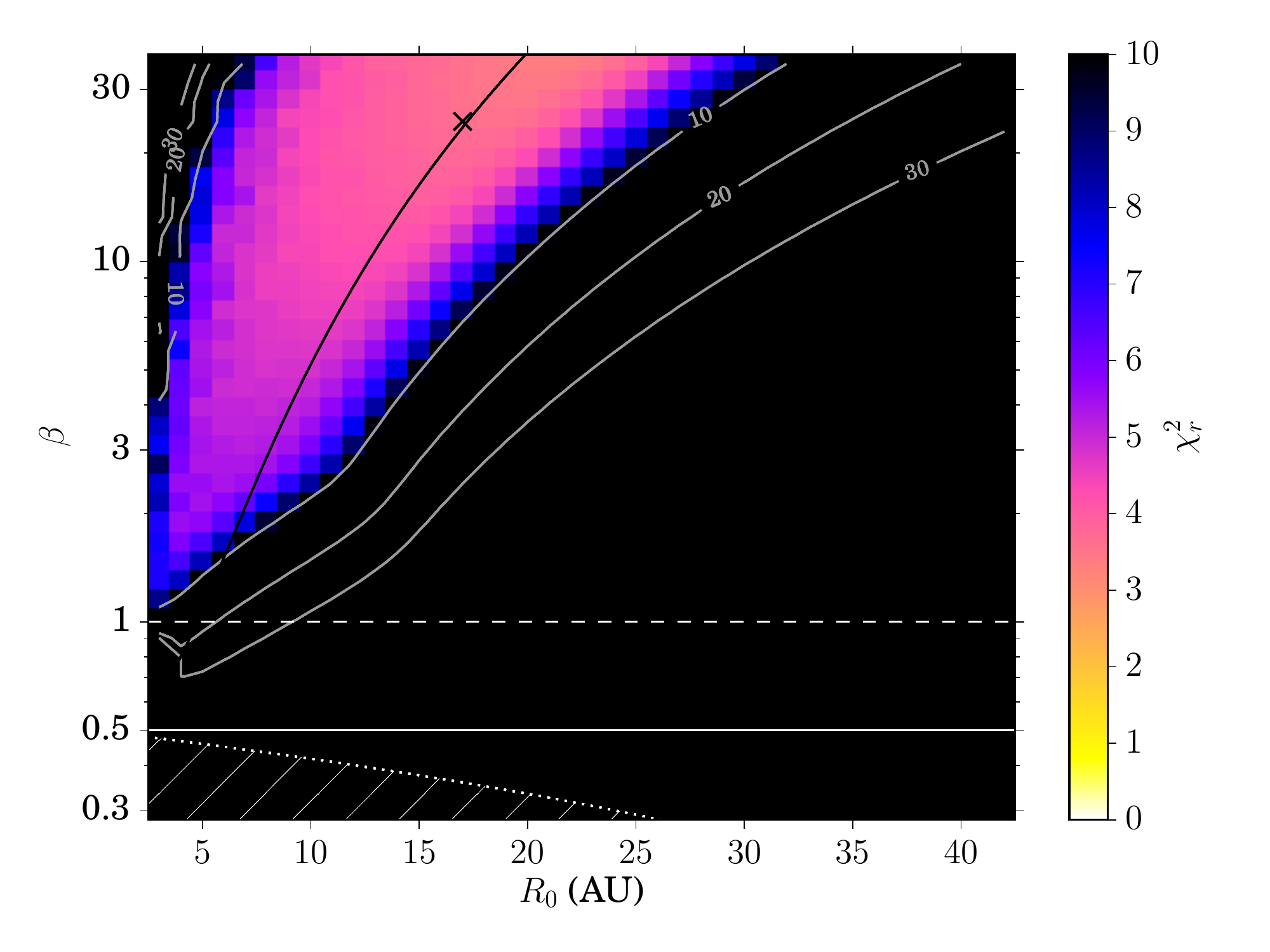}
\subcaption{Forward case}
}
\parbox{0.5\textwidth}{ \includegraphics[width=0.5\textwidth, trim=0 0.5cm 0 0.5cm,clip]{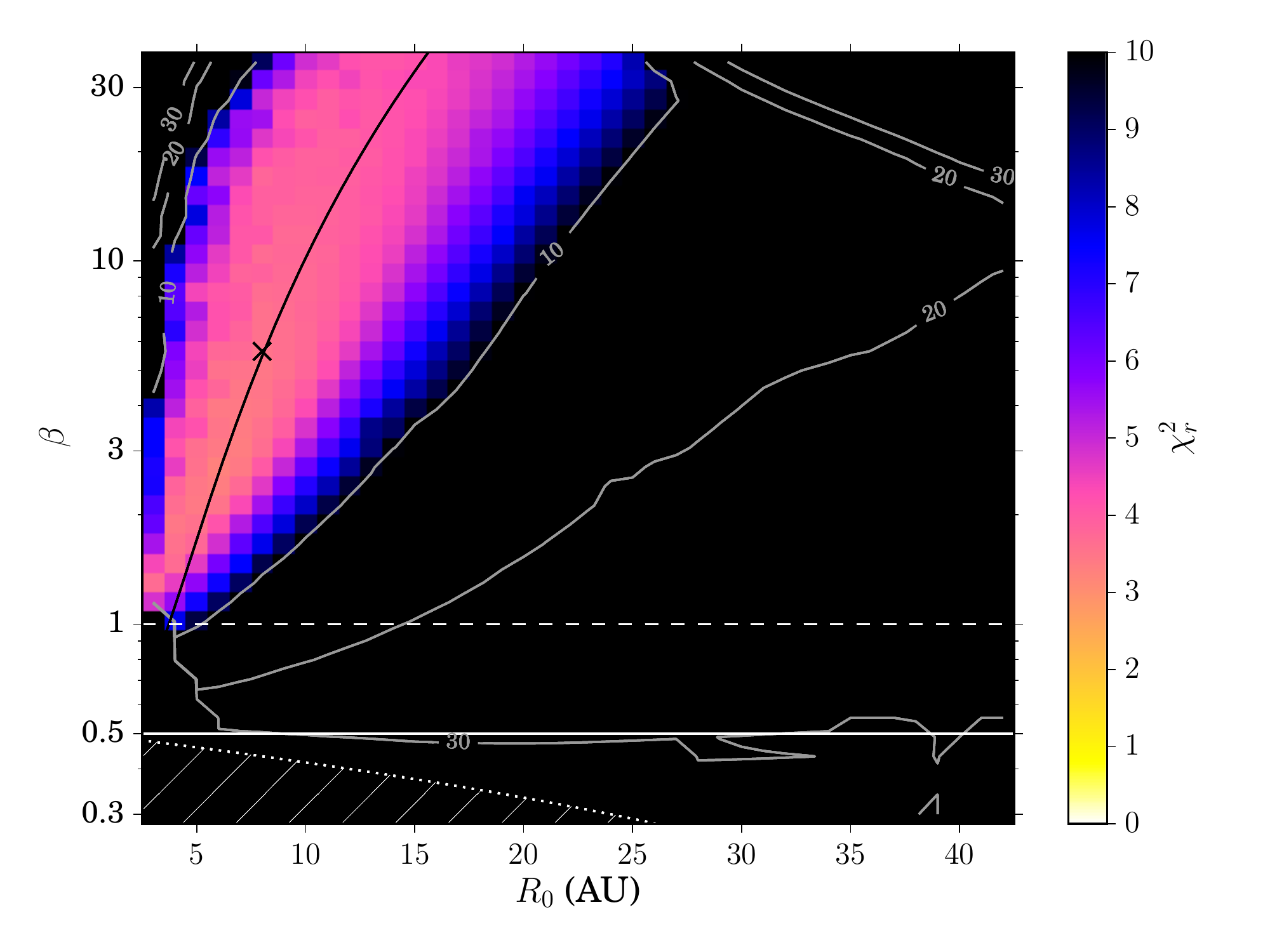}
\subcaption{Backward case}
}
}
\caption{\it Orbiting parent body. \rm 
\XX\ maps for the grouped release cases.}
\label{figMapGroup}
\end{figure*}

\end{appendix}


\end{document}

%% file: tabLatexStatic.tex
Structure A as a reference & 2011.6 $\pm$ 0.1 & 2005.0 $\pm$ 0.1 & 1996.2 $\pm$ 0.1 & 1990.3 $\pm$ 0.1 & 1982.2 $\pm$ 0.1 \\
Structure B as a reference & 2011.6 $\pm$ 0.1 & 2005.0 $\pm$ 0.1 & 1996.2 $\pm$ 0.1 & 1990.3 $\pm$ 0.1 & 1982.2 $\pm$ 0.1 \\
Structure C as a reference & 2011.6 $\pm$ 0.1 & 2005.0 $\pm$ 0.1 & 1996.1 $\pm$ 0.1 & 1990.2 $\pm$ 0.1 & 1982.1 $\pm$ 0.1 \\
Structure D as a reference & 2011.5 $\pm$ 0.1 & 2004.9 $\pm$ 0.1 & 1996.1 $\pm$ 0.1 & 1990.1 $\pm$ 0.1 & 1982.0 $\pm$ 0.1 \\
Structure E as a reference & 2011.4 $\pm$ 0.1 & 2004.8 $\pm$ 0.1 & 1996.0 $\pm$ 0.1 & 1990.0 $\pm$ 0.1 & 1981.9 $\pm$ 0.1 \\
\hline
Average & 2011.6 $\pm$ 0.1 & 2005.0 $\pm$ 0.1 & 1996.1 $\pm$ 0.1 & 1990.2 $\pm$ 0.1 & 1982.1 $\pm$ 0.1 \\ 
\hline

%% file: tabLatexFree.tex
Structure A as a reference & 2004.2 $\pm$ 0.6 & 2003.3 $\pm$ 0.9 & 1989.0 $\pm$ 0.7 & 1989.6 $\pm$ 0.7 & 1990.6 $\pm$ 0.7 \\
Structure B as a reference & 2003.7 $\pm$ 0.8 & 2002.9 $\pm$ 0.4 & 1988.4 $\pm$ 0.6 & 1989.0 $\pm$ 0.5 & 1990.0 $\pm$ 0.5 \\
Structure C as a reference & 2003.4 $\pm$ 0.6 & 2002.5 $\pm$ 0.6 & 1988.0 $\pm$ 0.3 & 1988.6 $\pm$ 0.1 & 1989.6 $\pm$ 0.2 \\
Structure D as a reference & 2004.9 $\pm$ 0.6 & 2003.9 $\pm$ 0.5 & 1989.8 $\pm$ 0.1 & 1990.5 $\pm$ 0.2 & 1991.6 $\pm$ 0.1 \\
Structure E as a reference & 2004.0 $\pm$ 0.6 & 2003.1 $\pm$ 0.5 & 1988.8 $\pm$ 0.2 & 1989.4 $\pm$ 0.2 & 1990.4 $\pm$ 0.2 \\
\hline
Average & 2004.1 $\pm$ 0.7 & 2003.1 $\pm$ 0.6 & 1989.1 $\pm$ 0.5 & 1989.3 $\pm$ 0.4 & 1990.6 $\pm$ 0.4 \\ 
\hline

%% file: tabLatexFront.tex
Structure A as a reference & 2000.8 $\pm$ 0.2 & 1999.4 $\pm$ 0.1 & 1997.2 $\pm$ 0.1 & 1995.8 $\pm$ 0.2 & 1994.1 $\pm$ 0.2 \\
Structure B as a reference & 2000.2 $\pm$ 0.1 & 1998.9 $\pm$ 0.1 & 1996.8 $\pm$ 0.1 & 1995.4 $\pm$ 0.1 & 1993.7 $\pm$ 0.2 \\
Structure C as a reference & 2000.4 $\pm$ 0.1 & 1999.1 $\pm$ 0.1 & 1996.9 $\pm$ 0.1 & 1995.5 $\pm$ 0.1 & 1993.8 $\pm$ 0.1 \\
Structure D as a reference & 2000.4 $\pm$ 0.2 & 1999.1 $\pm$ 0.1 & 1996.9 $\pm$ 0.1 & 1995.5 $\pm$ 0.1 & 1993.8 $\pm$ 0.1 \\
Structure E as a reference & 2000.3 $\pm$ 0.2 & 1999.0 $\pm$ 0.2 & 1996.9 $\pm$ 0.1 & 1995.5 $\pm$ 0.1 & 1993.8 $\pm$ 0.1 \\
\hline
Average & 2000.4 $\pm$ 0.2 & 1999.2 $\pm$ 0.1 & 1996.9 $\pm$ 0.1 & 1995.5 $\pm$ 0.1 & 1993.8 $\pm$ 0.1 \\ 
\hline

%% file: tabLatexBack.tex
Structure A as a reference & 1990.7 $\pm$ 0.2 & 1991.1 $\pm$ 0.1 & 1992.1 $\pm$ 0.1 & 1993.0 $\pm$ 0.1 & 1994.5 $\pm$ 0.2 \\
Structure B as a reference & 1990.0 $\pm$ 0.1 & 1990.4 $\pm$ 0.1 & 1991.3 $\pm$ 0.1 & 1992.2 $\pm$ 0.1 & 1993.5 $\pm$ 0.1 \\
Structure C as a reference & 1989.7 $\pm$ 0.1 & 1990.1 $\pm$ 0.1 & 1991.0 $\pm$ 0.1 & 1991.8 $\pm$ 0.1 & 1993.1 $\pm$ 0.1 \\
Structure D as a reference & 1990.0 $\pm$ 0.1 & 1990.4 $\pm$ 0.1 & 1991.3 $\pm$ 0.1 & 1992.2 $\pm$ 0.1 & 1993.6 $\pm$ 0.1 \\
Structure E as a reference & 1989.6 $\pm$ 0.2 & 1990.0 $\pm$ 0.1 & 1990.8 $\pm$ 0.1 & 1991.6 $\pm$ 0.1 & 1993.0 $\pm$ 0.1 \\
\hline
Average & 1990.0 $\pm$ 0.2 & 1990.7 $\pm$ 0.2 & 1991.4 $\pm$ 0.2 & 1992.1 $\pm$ 0.2 & 1993.5 $\pm$ 0.3 \\ 
\hline